\documentclass[journal]{IEEEtran}

\usepackage{cite,amssymb,color,amsmath,graphicx}

\newtheorem{secthm}{Theorem}[section]
\newtheorem{seccor}[secthm]{Corollary}

\newtheorem{secprop}[secthm]{Proposition}
\newtheorem{secdefn}[secthm]{Definition}
\newtheorem{secrem}[secthm]{Remark}
\newtheorem{secasm}[secthm]{Assumption}
\newtheorem{secsasm}[secthm]{Standing Assumption}

\newcommand{\bE} { {\mathbb E}}

\newcommand{\bP} { {\mathbb P}}
\newcommand{\bR} { {\mathbb R}}
\newcommand{\bS} { {\mathbb S}}
\newcommand{\bZ} { {\mathbb Z}}

\newcommand{\cB} { {\cal B}}

\newcommand{\cF} { {\cal F}}

\newcommand{\cL} { {\cal L}}
\newcommand{\cM} { {\cal M}}
\newcommand{\cS} { {\cal S}}
\newcommand{\cT} { {\cal T}}

\newcommand{\cX} { {\cal X}}

\allowdisplaybreaks[4]
\def\red{\hfill $\lhd$}

\begin{document}
\title{Contraction Analysis of Discrete-time Stochastic Systems}
\author{Yu Kawano, {\it Member, IEEE}, Yohei Hosoe, {\it Member, IEEE}
\thanks{Y.~Kawano is with the Graduate School of Advanced Science and Engineering, Hiroshima University, Higashi-Hiroshima, Japan (email: ykawano@hiroshima-u.ac.jp).}
\thanks{Y.~Hosoe is with the Department of Electrical Engineering, Kyoto University, Nishikyo-ku, Kyoto 615-8510, Japan (email: hosoe@kuee.kyoto-u.ac.jp).}
\thanks{This work of Kawano was supported in part by JSPS KAKENHI Grant Numbers JP21K14185 and JP21H04875. This work of Hosoe was supported in part by JSPS KAKENHI Grant Numbers JP20K04546.}}
\maketitle

\IEEEpeerreviewmaketitle

\begin{abstract}
In this paper, we develop a novel contraction framework for stability analysis of discrete-time nonlinear systems with parameters following stochastic processes. For general stochastic processes, we first provide a sufficient condition for uniform incremental exponential stability (UIES) in the first moment with respect to a Riemannian metric. Then, focusing on the Euclidean distance, we present a necessary and sufficient condition for UIES in the second moment. By virtue of studying general stochastic processes, we can readily derive UIES conditions for special classes of processes, e.g., i.i.d. processes and Markov processes, which is demonstrated as selected applications of our results. 
\end{abstract}

\section{Introduction}
Starting with a seminal paper~\cite{LS:98}, contraction theory draws attention from the systems and control community as a new differential geometric framework for stability analysis of nonlinear systems. Differently from the standard Lyapunov analysis of an equilibrium point (e.g., \cite{Khalil:96,HC:08}), incremental stability (i.e., stability of a pair of trajectories) \cite{Angeli:02} is studied by lifting the Lyapunov function to the tangent bundle~\cite{FS:14}. Revisiting nonlinear control theory from this new angle has resulted in so-called differential approaches to, for instance, control design~\cite{RSJ:17,RSJ:20,MS:17,KO:17}, observer design~\cite{BS:15,AJP:16,DCH:14}, dissipativity theory~\cite{Schaft:13,FS:13,KCS:21}, and balancing theory~\cite{KS:15,KS:17}. Along with them, contraction (stability) analysis itself is in the middle of development in various problem settings; see, e.g., \cite{FS:15,KBC:20} for monotone systems, e.g.,~\cite{BLR:14,FHB:16} for switched systems, e.g., \cite{PTS:09,TCS:20} for systems under stochastic input noise, and \cite{LXL:20} for stochastic switched impulsive systems, a kind of Markov jump systems.

In this paper, we aim at newly developing contraction theory for discrete-time nonlinear systems with parameters following stochastic processes. None of the aforementioned papers deals with this class of systems; most of them focus on continuous-time deterministic systems. The aforementioned papers \cite{PTS:09,TCS:20,LXL:20} studying stochastic systems are for continuous-time systems. In the discrete-time case, \cite{LS:98,TRK:18} and \cite{PS:13,TC:19} have studied deterministic systems and systems under stochastic input noise, respectively. Other than input noise, randomness is not incorporated in contraction analysis of discrete-time systems. In other words, there is no contraction framework to analyze discrete-time systems with random parameters such as Markov jump systems~\cite{CFM:05} and the systems with white parameters~\cite{Koning:92}. This is in contrast to a massive amount of researches on discrete-time Markov jump linear/nonlinear systems in the history, e.g., \cite{CFM:05,lLD:06,GC:13,ZHZ:14} and recent rapid increase in the number of researches for machine learning to construct stochastic models from discrete-time empirical data, e.g.,~\cite{SLA:12,BCV:13,SSW:15,GBC:16}. When studying stochastic systems, typically we specify the class of stochastic processes into, for instance, i.i.d. and Markovian, which can be viewed as ad hoc approaches because depending on processes, different stability conditions are obtained. For developing unified theory to deal with each process simultaneously, recently the paper~\cite{HH:22} gives second moment stability conditions for general stochastic processes in the discrete-time linear case, which contains the existing conditions for i.i.d.\cite{OM:13,HH:19} and Markovian \cite{CFM:05,CF:14} as special cases. 

Inspired by~\cite{HH:22}, in this paper, we deal with general stochastic processes. To begin with contraction analysis of discrete-time nonlinear stochastic systems, we introduce a new stability notion, uniform incremental exponential stability (UIES) in the $p$th moment with respect to the Riemannian metric, which reduces to the standard $p$th moment stability \cite{Kozin:69,FLF:94} when the distance is Euclidean, and a trajectory is fixed on an equilibrium point. As the first main result of this paper, we provide a sufficient condition for UIES in the first moment. Then, as the second main contribution, focusing on the Euclidean distance, we present a necessary and sufficient condition for UIES in the second moment; second moment stability is stronger than first moment stability. By virtue of developing unified theory for general stochastic processes, we show that specifying processes readily yields UIES conditions for i.i.d. processes or Markov processes. Even UIES conditions in each specialized case are new contributions of this paper on their own, due to lack of contraction theory for discrete-time stochastic systems.

The remainder of this paper is organized as follows. To understand the whole picture of this paper, Section~\ref{Det:sec} summarizes contraction analysis of discrete-time deterministic systems with respect to the Euclidean distance \cite{TRK:18} and then extends this to a Riemannian metric. Section~\ref{PF:sec} shows the discrete-time stochastic systems considered in this paper and provides the notion of UIES in the $p$th moment. Section~\ref{ISC:sec} presents the UIES conditions for general stochastic processes, and these conditions are applied to i.i.d. processes and Markov processes in Section~\ref{App:sec}. Some of the proposed stability conditions are applied to stabilizing controller design of a mechanical system with a random parameter and observer design for a Markov jump system in Section~\ref{Ex:sec}. Concluding remarks are given in Section~\ref{Con:sec}. All proofs are presented in the Appendix.

{\it Notation}
The sets of real numbers and integers are denoted by~$\bR$ and~$\bZ$, respectively. Subsets of~$\bZ$ are defined by~$\bZ_{k_0+}:=\bZ \cap [k_0, \infty)$ and~$\bZ_{k_0-}:=\bZ \cap (-\infty, k_0]$ for $k_0 \in \bZ$. Another subset of~$\bZ$ is defined by~$\bZ_{[k_0,k]}:=\bZ \cap [k_0, k]$ for $k_0 \in \bZ$ and $k \in \bZ_{k_0+}$, where $\bZ_{[k_0,k_0]}:=\{k_0\}$.
The identity matrix is denoted by~$I$ irrespective of its size. The set of $n\times n$ symmetric matrices is denoted by~$\bS^{n\times n}$, and that of symmetric and positive (resp. semi) definite matrices is denoted by~$\bS_{\succ 0}^{n\times n}$ (resp. $\bS_{\succeq 0}^{n\times n}$). For~$P, Q \in \bR^{n \times n}$,  $P \succ Q$ (resp. $P \succeq Q$) means~$P - Q \in \bS_{\succ 0}^{n\times n}$ (resp. $P - Q \in \bS_{\succeq 0}^{n\times n}$). The Euclidean norm of a vector~$x \in \bR^n$ is denoted by~$|x|$.

Let $(\Omega, \cF, \bP)$ be a complete probability space, where~$\Omega$,~$\cF$, and~$\bP$ denote a sample space, $\sigma$-algebra, and probability measure, respectively. For the sake of notational simplicity, an~$\cX$-valued random variable~$\xi_0: (\Omega, \cF) \to (\cX, \cB(\cX))$ is described by~$\xi_0: \Omega \to \cX$, where~$\cB(\cX)$ denotes the Borel $\sigma$-algebra on~$\cX$. An~$X^\cT$-valued stochastic process~$\xi:=(\xi_k)_{k \in \cT}$ on $\cT \subset \bZ$ is defined as a mapping~$\xi: \Omega (\times \cF) \to \cX^\cT (\times \cB ( \cX^\cT ))$. For some~$k_0 \in \bZ$, a subsequence of a stochastic process~$\xi:\Omega \to \cX^\bZ$ is denoted by~$\xi^{k_0+}= (\xi_k)_{k \in \bZ_{k_0+}} : \Omega \to \cX^{\bZ_{k_0+}}$, and $\xi^{k_0-}$ is defined similarly. The support of $\xi^{(k_0-1)-}$ is denoted by~$\hat \Xi^{(k_0-1)-}$.  When~$\xi^{(k_0-1)-} = \hat \xi^{(k_0-1)-} \in \hat \Xi^{(k_0-1)-}$, the conditional expectation of a function of~$\xi^{k_0+}$ given~$\xi^{(k_0-1)-}$ is denoted by~$\bE_0 [(\cdot)] := \bE [(\cdot) | \xi^{(k_0-1)-} = \hat \xi^{(k_0-1)-} ]$. Let~$\cF_k$ be the~$\sigma$-algebra generated by a subsequence~$\xi_{k_0},\xi_{k_0+1},\dots,\xi_k$ of a stochastic process~$\xi$ under the initial condition~$\xi^{(k_0-1)-} = \hat \xi^{(k_0-1)-} \in \hat \Xi^{(k_0-1)-}$. Then, $(\cF_k)_{k \in \bZ_{k_0+}}$ is a filtration on $(\Omega, \cF, \bP)$ for each~$\hat \xi^{(k_0-1)-} \in \hat \Xi^{(k_0-1)-}$, namely an increasing family $(\cF_k)_{k \in \bZ_{k_0+}}$ of sub-$\sigma$-algebras of~$\cF$. The conditional expectation of a function of~$\xi^{k_0+}$ given~$\cF_k$ is denoted by~$\bE_0 [(\cdot)| \cF_k]$. This conditional expectation satisfies~$\bE_0 [\bE_0 [(\cdot)| \cF_{k_2}] | \cF_{k_1}] = \bE_0 [(\cdot)| \cF_{k_1}]$ for each~$k_1 \in \bZ_{k_0+}$ and every~$k_2 \in \bZ_{k_1+}$.

\section{Reviews and Generalizations of Results for Deterministic Systems}\label{Det:sec}
To understand the whole picture of this paper, we first review results on contraction analysis of discrete-time nonlinear deterministic systems~\cite{TRK:18, LS:98}. In these literature, the Euclidean distance is used as a metric. In this paper, we show that some sufficiency result can be extended to a Riemannian metric as for continuous-time systems~\cite{FS:14}.

Consider the following nonlinear deterministic system:
\begin{align}
z_{k+1} = g_k( z_k), \; k \in \bZ,
\label{sys_dtm}
\end{align}
where~$g_k: \bR^n \to \bR^n$ is of class $C^1$ for each~$k \in \bZ$. Note that~$\bR^n$ is positively invariant. For the sake of notational simplicity, let~$\psi_k(k_0,z_{k_0})$ denote the solution to the system~\eqref{sys_dtm} at~$k \in \bZ_{k_0+}$ under the initial condition~$(k_0, z_{k_0}) \in \bZ \times \bR^n$. Namely,
\begin{align}
&z_k = \psi_k(k_0, z_{k_0}), \; k \in \bZ_{k_0+}, \nonumber\\
&\psi_{k+1} (k_0, z_{k_0}) = g_k( \psi_k(k_0, z_{k_0})), \; k \in \bZ_{k_0+} \label{sys_dtm_sol}
\end{align}
for each~$(k_0, z_{k_0}) \in \bZ \times \bR^n$, where~$\psi_{k_0}(k_0, z_{k_0}) = z_{k_0}$.

As will be clear later, we use the following variational system of~\eqref{sys_dtm} along~$\psi_k (k_0, z_{k_0})$ in contraction analysis:
\begin{align}
\delta z_{k+1} = \frac{\partial g_k (\psi_k(k_0, z_{k_0}))}{\partial \psi_k} \delta z_k.
\label{vsys_dtm}
\end{align}
Using variational systems,  incremental stability conditions have been developed; this stability notion is defined as follows.
\begin{secdefn}
Let $d:\bR^n \times \bR^n \to \bR$ be a distance\footnote{A function $d:\bR^n \times \bR^n \to \bR$ is said to be distance if 1) $d(\cdot,\cdot) \ge 0$, 2) $d(x,x') = 0$ if and only if $x=x'$ for all~$x,x' \in \bR^n$, and 3) $d(x, x'') \le d(x,x') + d(x', x'')$ for all~$x, x', x'' \in \bR^n$.}. The system~\eqref{sys_dtm} is said to be {\it uniformly incrementally exponentially stable} (UIES) (with respect to~$d$) if there exist $a>0$ and $\lambda \in (0,1)$ such that
\begin{align*}
d(z'_k, z''_k) \le a \lambda^{k-k_0} d (z'_{k_0}, z''_{k_0}), \; \forall k \in \bZ_{k_0+}
\end{align*}
for each~$(k_0, (z'_{k_0}, z''_{k_0})) \in \bZ \times (\bR^n \times \bR^n)$.
\red
\end{secdefn}

When the distance is Euclidean, the following condition for UIES has been derived~\cite[Theorem 15]{TRK:18}, where the condition below is slightly different from the original one, but is equivalent to it.
\begin{secprop}\label{IES_dtm:prop}
A system~\eqref{sys_dtm} is UIES with respect to the Euclidean distance if and only if there exist~$c_1, c_2 > 0$, $\lambda \in (0,1)$, and~$P: \bZ \times \bR^n \to \bS_{\succ 0}^{n\times n}$ such that
\begin{align}
&c_1^2 I \preceq P(k_0, z_{k_0}) \preceq c_2^2 I, \label{Pbd_dtm:cond}\\
&\frac{\partial^\top g_{k_0} (z_{k_0})}{\partial z} P(k_0+1, g_{k_0} (z_{k_0})) \frac{\partial g_{k_0} (z_{k_0})}{\partial z}
\preceq \lambda^2 P(k_0, z_{k_0})
\label{IES_dtm:cond}
\end{align}
for all~$(k_0, z_{k_0}) \in \bZ \times \bR^n$.
\red
\end{secprop}

Inspired by results for continuous-time systems~\cite{FS:14}, we generalize the condition~\eqref{Pbd_dtm:cond} to study UIES with respect to a more general Riemannian metric than Euclidean as stated below. This result can be hypothesized from Proposition~\ref{IES_dtm:prop}, but has not been proven before. More importantly, its proof gives an insight into analysis of stochastic systems, the main interests of this paper. Thus, the proof is also provided in Appendix~\ref{IES_dtm:app}.
\begin{secthm}\label{IES_dtm:thm}
A system~\eqref{sys_dtm} is UIES if there exist~$c_1, c_2 > 0$, $\lambda \in (0,1)$, $\hat P: \bR^n \to \bS_{\succ 0}^{n\times n}$ of class $C^1$, and $P: \bZ \times \bR^n \to \bS_{\succ 0}^{n\times n}$ such that
\begin{align}
c_1^2 \hat P(z_{k_0}) \preceq P(k_0, z_{k_0}) \preceq c_2^2 \hat P(z_{k_0}) \label{Pbd_dtm:cond2}
\end{align}
and~\eqref{IES_dtm:cond} hold for all~$(k_0, z_{k_0}) \in \bZ \times \bR^n$.
\red
\end{secthm}

The objective of this paper is to extend Proposition~\ref{IES_dtm:prop} and Theorem~\ref{IES_dtm:thm} to stochastic systems introduced in the next section. 

\section{Problem Formulations}\label{PF:sec}
Hereafter, we focus on the stochastic systems stated in this section. Let~$\xi := (\xi_k)_{k \in \bZ}: \Omega \to (\bR^m)^\bZ$ be a stochastic process. Differently from usual analysis, we consider general~$\xi$, i.e., do not focus on specific $\xi$ such as i.i.d. or Markovian. Throughout this paper, we assume that a vector-valued function~$f_k:\bR^n \times \bR^m \to \bR^n$, $k \in \bZ$ defining the system dynamics satisfies the following assumption.
\begin{secsasm}\label{f:asm}
At each~$k \in \bZ$, $f_k (x, \eta )$ is semi-differentiable with respect to~$x$. Moreover, $f_k$ and its semi-differentiation\footnote{In general, $\partial f_k/\partial x$ denotes the partial derivative of~$f_k$ with respect to~$x$, but we use this to denote a semi-differentiation by abuse of notation.}, denoted by $\partial f_k/\partial x$, are both piecewise continuous on~$\bR^n \times \bR^m$ at each~$k \in \bZ$.
\red
\end{secsasm}
\begin{secrem}
Since the piecewise continuous function is Borel measurable~\cite[Problem 2.2]{AL:06}, Assumption~\ref{f:asm} implies the following:
\begin{enumerate}
\item at each~$k \in \bZ$, $f_k: (\bR^n \times \bR^m, \cB(\bR^n \times \bR^m)) \to (\bR^n, \cB(\bR^n))$ is Borel measurable;
\item at each~$k \in \bZ$, $\partial f_k/\partial x: (\bR^n \times \bR^m, \cB(\bR^n \times \bR^m)) \to (\bR^{n \times n}, \cB(\bR^{n \times n}))$ is Borel measurable. 
\end{enumerate}
Note that the piecewise continuity of $\partial f_k/\partial x$ does not imply that of $f_k$ in general.
\red
\end{secrem}

\begin{secrem}\label{switch:rem}
An example of~$f_k$ satisfying Assumption~\ref{f:asm} is of piecewise~$C^1$. The set of $f_k$ contains $\bar f_k (x, \eta, s)$ with a switching function $s(x,\eta)$. Let~$\cS_i \subset \bR^n \times \bR^m$, $i \in \cM :=\{ 1,\dots,M\}$ and $s:\bR^n \times \bR^m \to \cM$, respectively, denote a finite family of disjoint subsets and a switching function such that $\cup_{i \in \cM} \cS_i = \bR^n \times \bR^m$, and the semi-differentiation $\partial s/\partial x$ is well defined as the zero function for each $\cS_i$, $i \in \cM$ and thus on $\bR^n \times \bR^m$. If $\bar f_k(x, \eta, s)$ is semi-differentiable with respect to $x$ and $s$, and if $\bar f_k$ and its semi-differentiations are piecewise continuous, then $\bar f_k (x, \eta, s(x,\eta))$ satisfies Assumption~\ref{f:asm}. Therefore, $f_k$ can also be used to describe switched systems.
\red
\end{secrem}

Our interest in this paper is the following discrete-time nonlinear stochastic system:
\begin{align}
x_{k+1} = f_k(x_k, \xi_k), \; k \in \bZ_{k_0+}
\label{sys}
\end{align}
for a given deterministic initial condition~$(k_0, x_{k_0}, \hat \xi^{(k_0-1)-}) \in \bZ \times \bR^n  \times \hat \Xi^{(k_0-1)-}$; $k_0$, $x_{k_0}$, and $\hat \xi^{(k_0-1)-}$ denote the initial time, the initial state of the system~\eqref{sys}, and  the initial state of the stochastic process~$\xi$, respectively. To emphasize that $(x_k)_{k \in \bZ_{k_0+}}: \Omega \to (\bR^n)^{\bZ_{k_0+}}$ is a stochastic process under the initial condition~$(k_0, x_{k_0}, \hat \xi^{(k_0-1)-}) \in \bZ \times \bR^n \times \hat \Xi^{(k_0-1)-}$, this is also denoted by $(\phi_k(\xi^{(k-1)-}; k_0, x_{k_0}, \hat \xi^{(k_0-1)-}))_{k \in \bZ_{k_0+}}$ or simply $(\phi_k(\xi^{(k-1)-}))_{k \in \bZ_{k_0+}}$. Namely, it follows that
\begin{align}
&x_k=\phi_k(\xi^{(k-1)-}),\label{sys_sol0}\\
&\phi_{k+1}(\xi^{k-}) = f_k(\phi_k(\xi^{(k-1)-}), \xi_k), \; k \in \bZ_{k_0+} \label{sys_sol}
\end{align}
for each $(k_0, x_{k_0}, \hat \xi^{(k_0-1)-}) \in \bZ \times \bR^n  \times \hat \Xi^{(k_0-1)-}$, where~$\phi_{k_0} = x_{k_0}$.

As for contraction analysis of deterministic systems, we use the following variational system of~\eqref{sys} along $\phi_k(\xi^{(k-1)-}; k_0, x_{k_0}, \hat \xi^{(k_0-1)-})$:
\begin{align}
\delta x_{k+1} = \frac{\partial f_k(\phi_k (\xi^{(k-1)-}), \xi_k)}{\partial \phi_k} \delta x_k, \; k \in \bZ_{k_0+}
\label{vsys}
\end{align}
for the same deterministic initial time~$k_0 \in \bZ$ as the system~\eqref{sys} and a given deterministic initial state~$\delta x_{k_0} \in \bR^n$ of the variational system. From its definition, the variational system is also a stochastic system. 

Next, we introduce the notion of incremental stability to the stochastic systems as follows.
\begin{secdefn}
Consider the system~\eqref{sys}, and let $d:\bR^n \times \bR^n \to \bR$ be a distance. Then, the system is said to be {\it UIES in the $p$th moment} (with respect to $d$) if there exist $a>0$ and $\lambda \in (0,1)$ such that
\begin{align}
\bE_0 [d^p(x'_k,x''_k) ] \le a^p \lambda^{p(k-k_0)} d^p(x'_{k_0}, x''_{k_0}), \; \forall k \in \bZ_{k_0+}
\label{IES:def}
\end{align}
for each~$(k_0, (x'_{k_0}, x''_{k_0}), \hat \xi^{(k_0-1)-}) \in \bZ \times (\bR^n \times \bR^n) \times \hat \Xi^{(k_0-1)-}$. 
\red
\end{secdefn}

The above stability notion reduces to the standard moment stability if we choose $d$ as the Euclidean distance, and the origin is an equilibrium point, i.e., $f_k(0,\xi_k) \equiv 0$, $k \in \bZ$. In fact, for $(x'_{k_0}, x''_{k_0}) =(x_{k_0}, 0)$ and the Euclidean distance, \eqref{IES:def} becomes
\begin{align*}
\bE_0 [|x_k|^p ] \le a^p \lambda^{p(k-k_0)} |x_{k_0}|^p, \; \forall k \in \bZ_{k_0+}.
\end{align*}
Especially for $p=2$, this property is also called mean square stability \cite{Kozin:69,FLF:94}. However, for~$p=1$, this is different from mean stability \cite{FLF:94} studying $\bE_0 [x_k]$ because the Euclidean norm $| \cdot |$ is taken. 

\section{Incremental Stability Conditions}\label{ISC:sec}
\subsection{With respect to Riemannian Metrics}
In this section, inspired by results for linear stochastic systems~\cite{HH:22}, we consider extending Proposition~\ref{IES_dtm:prop} and Theorem~\ref{IES_dtm:thm} to the stochastic systems~\eqref{sys}. Since the sufficiency of Proposition~\ref{IES_dtm:prop} is a special case of Theorem~\ref{IES_dtm:thm}, we first focus on deriving the counterpart of Theorem~\ref{IES_dtm:thm}. The main difference from the deterministic case is that we consider~$P$ depending on the stochastic process~$\xi$. To make the arguments of~$P$ clear, let us introduce the time shift operator~$S_k:(\bR^m)^{\bZ_{k+}} \to (\bR^m)^{\bZ_{0+}}$ for processes such that~$\zeta^{0+} = S_k \xi^{k+}$ is defined by~$\zeta_0 = \xi_k$, $\zeta_1 = \xi_{k+1},\dots$, where $\zeta^{0+} = S_k \xi^{k+}$ is $\cF_k$-measurable. Now, we are ready to present the first main result of this paper.
\begin{secthm}\label{IES:thm}
A system~\eqref{sys} is UIES in the first moment if there exist $c_1, c_2 > 0$, $\lambda \in (0,1)$, $\hat P: \bR^n \to \bS_{\succ 0}^{n\times n}$ of class~$C^1$, and~$P: \bZ \times \bR^n \times (\bR^m)^{\bZ_{0+}} \to [-\infty, \infty]^{n \times n}$  such that
\begin{align}
&c_1^2 \hat P(x_{k_0}) \preceq \bE_0 [P(k_0, x_{k_0}, S_{k_0} \xi^{k_0+}) ] \preceq c_2^2 \hat P(x_{k_0}), \label{Pbd:cond}\\
&\bE_0 \left[ \frac{\partial^\top f_{k_0}(x_{k_0},\xi_{k_0})}{\partial x_{k_0}} \right. \nonumber\\
&\hspace{7mm} \bE_0 [ P(k_0+1, f_{k_0}(x_{k_0}, \xi_{k_0}), S_{k_0+1} \xi^{(k_0+1)+})| \cF_{k_0}] \nonumber\\
&\hspace{5mm}\left. \frac{\partial f_{k_0}(x_{k_0},\xi_{k_0})}{\partial x_{k_0}} \right]
\preceq \lambda^2 \bE_0[ P(k_0, x_{k_0}, S_{k_0} \xi^{k_0+}) ]
\label{IES:cond}
\end{align}
for all~$(k_0, x_{k_0}, \hat \xi^{(k_0-1)-} ) \in \bZ \times \bR^n \times \hat \Xi^{(k_0-1)-}$. 
\red
\end{secthm}

\begin{secrem}\label{IES:rem}
From the proof of Theorem~\ref{IES:thm} in Appendix~\ref{IES:app}, one notices that a (non-uniform) IES condition in the first moment can readily be obtained by replacing $c_1$, $c_2$, $\lambda$, and $\hat P$ with those depending on~$k_0$. By IES in the $p$th moment at~$k_0 \in \bZ$, we mean that  there exist $a(k_0)>0$ and $\lambda(k_0) \in (0,1)$ such that
\begin{align*}
\bE_0 [d^p(x'_k, x''_k) ] \le a^p(k_0) \lambda^{p(k-k_0)} (k_0) d^p(x'_{k_0}, x''_{k_0}), \nonumber\\\
 \forall k \in \bZ_{k_0+}
\end{align*}
for each~$((x'_{k_0}, x''_{k_0}), \hat \xi^{(k_0-1)-}) \in (\bR^n \times \bR^n) \times \hat \Xi^{(k_0-1)-}$. 
On the other hand, Theorem~\ref{IES:thm} can be generalized to incremental stability analysis on an open subset~$D \subset \bR^n$ when~$f_k : D \times \bR^m \to D$, $k \in \bZ$ because $D$ is a (robustly) positively invariant set for such~$f_k$.
\red
\end{secrem}

Theorem~\ref{IES:thm} reduces to Theorem~\ref{IES_dtm:thm} for the deterministic systems. This can be confirmed by considering a $\xi_k$-independent vector field $f_k(x_k,\xi_k)= g_k (x_k)$. In this case, we can take a $\xi_k$-independent matrix-valued function $P$. 

In Theorem~\ref{IES:thm}, we do not restrict the class of stochastic processes~$\xi$ into specific ones. Therefore, our framework can handle a variety of systems such as stochastic switching systems mentioned in Remark~\ref{switch:rem} by specifying properties of $\xi$ or further the structure of $f_k$ depending on problems. Utility of studying general $\xi$ is illustrated in Section~\ref{App:sec} by showing that restricting $\xi$ into a specific process readily derives UIES conditions for each process. 

\subsection{With respect to Euclidean Distances}
Theorem~\ref{IES:thm} provides a UIES condition with respect to a general distance. In this subsection, we focus on the Euclidean distance, which corresponds to specifying~$\hat P$ in Theorem~\ref{IES:thm} into the identity matrix. In this case, it is possible to obtain a UIES condition for second moment stability, stronger than first moment stability because we can avoid to apply the Cauchy--Schwarz inequality in contrast to the general Rimmanian metric case; for more details, see the proofs in Appendices~\ref{IES:app} and~\ref{IES_Euclid:app}. Moreover, we also have the converse proof. This can be viewed as a generalization of Proposition~\ref{IES_dtm:prop} to the general stochastic system~\eqref{sys}.
\begin{secthm}\label{IES_Euclid:thm}
A system~\eqref{sys} is UIES in the second moment with respect to the Euclidean distance if and only if there exist $c_1, c_2 > 0$, $\lambda \in (0,1)$, and~$P: \bZ \times \bR^n \times (\bR^m)^{\bZ_{0+}} \to [-\infty, \infty]^{n \times n}$ such that
\begin{align}
c_1^2 I \preceq \bE_0 [P(k_0, x_{k_0}, S_{k_0} \xi^{k_0+}) ] \preceq c_2^2 I \label{Pbd_Euclid:cond}
\end{align}
and \eqref{IES:cond} hold for all~$(k_0, x_{k_0}, \hat \xi^{(k_0-1)-} ) \in \bZ \times \bR^n \times \hat \Xi^{(k_0-1)-}$. 
\red
\end{secthm}

\begin{secrem}
A similar remark as Remark~\ref{IES:rem} holds. That is, a necessary and sufficient condition for IES in the second moment with respect to the Euclidean distance at~$k_0 \in \bZ$ can readily be derived based on Theorem~\ref{IES_Euclid:thm} by replacing $c_1, c_2 > 0$ and $\lambda \in (0,1)$ with those depending on~$k_0$.
\red
\end{secrem}

For UIES, the condition~\eqref{IES:cond} depends on the convergence rate~$\lambda \in (0, 1)$. As in the linear case~\cite[Lemma 3]{HH:22}, we can derive an alternative condition not depending on~$\lambda$. The proof is similar, and thus is omitted.
\begin{seccor}\label{IES_Euclid:cor}
Suppose that there exist $c_1, c_2 > 0$ and~$P: \bZ \times \bR^n \times (\bR^m)^{\bZ_{0+}} \to [-\infty, \infty]^{n \times n}$ such that~\eqref{Pbd_Euclid:cond} holds. Then, for some~$\lambda \in (0, 1)$, the condition~\eqref{IES:cond} holds for all~$(k_0, x_{k_0}, \hat \xi^{(k_0-1)-}) \in \bZ \times \bR^n \times \hat \Xi^{(k_0-1)-}$ if and only if there exists $c > 0$ such that
\begin{align*}
&\bE_0 \left[ \frac{\partial^\top f_{k_0}(x_{k_0},\xi_{k_0})}{\partial x_{k_0}} \right. \nonumber\\
&\hspace{7mm} \bE_0 [ P(k_0+1, f_{k_0}(x_{k_0}, \xi_{k_0}), S_{k_0+1} \xi^{(k_0+1)+})| \cF_{k_0}] \nonumber\\
&\hspace{5mm}\left. \frac{\partial f_{k_0}(x_{k_0},\xi_{k_0})}{\partial x_{k_0}} \right] \preceq  \bE_0[ P(k_0, x_{k_0}, S_{k_0} \xi^{k_0+}) ] - c^2 I
\end{align*}
for all~$(k_0, x_{k_0}, \hat \xi^{(k_0-1)-}) \in \bZ \times \bR^n \times \hat \Xi^{(k_0-1)-}$.
\red
\end{seccor}

In the linear case, Theorem~\ref{IES_Euclid:thm} reduces to \cite[Theorem 3]{HH:22}. Let $f_k(x_k, \xi_k)$ be linear, i.e., $f_k(x_k, \xi_k) = A(\xi_k) x_k$. Then, $P$ can be chosen to be independent of $x_k$. Therefore, \eqref{IES:cond} and \eqref{Pbd_Euclid:cond} reduce to
\begin{align*}
&c_1^2 I \preceq \bE_0 [P(S_{k_0} \xi^{k_0+}) ] \preceq c_2^2 I, \\
&\bE_0 \left[ A^\top (\xi_{k_0})  \bE_0 [ P(S_{k_0+1} \xi^{(k_0+1)+})| \cF_{k_0}] A (\xi_{k_0}) \right]\\
&\preceq \lambda^2 \bE_0[ P(S_{k_0} \xi^{k_0+}) ]
\end{align*}
for all~$(k_0, \hat \xi^{(k_0-1)-}) \in \bZ \times \hat \Xi^{(k_0-1)-}$. This is nothing but a necessary and sufficient condition for uniform exponential stability in the second moment for the general linear stochastic system in \cite[Theorem 3]{HH:22}.

\section{Applications}\label{App:sec}
In the previous section, we have presented incremental stability conditions for general stochastic systems~\eqref{sys}. In this section, we illustrate utility of the obtained conditions by applying them to specific classes of processes. In particular, we study cases where~$\xi$ follows temporally-independent processes or Markov processes. In most of literature of stochastic control, e.g.,~\cite{OM:13,GC:13,CF:14,ZHZ:14}, stability conditions have been separately developed for each special class of processes. By virtue of studying the general stochastic process~$\xi$, conditions for each special case are provided simply by restricting the class of~$\xi$ as in \cite{HH:22} about linear systems. Due to the lack of contraction analysis for stochastic systems, the obtained conditions in each special case are new contribution of this paper on their own. In this section, we only consider applying Theorem~\ref{IES_Euclid:thm}, but similar results corresponding to Theorem~\ref{IES:thm} and  Corollary~\ref{IES_Euclid:cor} as well as conditions for IES at $k_0 \in \bZ$ can readily be obtained.

\subsection{Temporally-Independent Processes}
In this subsection, we consider~$\xi$ satisfying the following assumption. Such~$\xi$ is called a \emph{temporally-independent process}.
\begin{secasm}\label{TI:asm}
For~$\xi = (\xi_k)_{k \in \bZ}$, the random vectors~$\xi_k$, $k \in \bZ$ are independently distributed.
\red
\end{secasm}

Under Assumption~\ref{TI:asm}, the conditions~\eqref{IES:cond} and~\eqref{Pbd_Euclid:cond} in Theorem~\ref{IES_Euclid:thm} are independent of~$\hat \xi^{(k_0-1)-}$ for each~$k_0 \in \bZ$. Hence, the conditional expectation can be replaced with the (standard) expectation. Then, by defining
\begin{align}
\hat P(k_0, x_{k_0}) :=\bE [P(k_0, x_{k_0}, S_{k_0} \xi^{k_0+}) ], \; (k_0, x_{k_0}) \in \bZ \times \bR^n,
\label{Phat}
\end{align}
we have the following corollary of Theorem~\ref{IES_Euclid:thm} without the proof. 

\begin{seccor}\label{IES_TI_Euclid:cor}
Suppose that Assumption~\ref{TI:asm} holds. A system~\eqref{sys} is UIES in the second moment with respect to the Euclidean distance if and only if there exist $c_1, c_2 > 0$, $\lambda \in (0,1)$, and $\hat P: \bZ \times \bR^n \to \bS_{\succ 0}^{n \times n}$ such that
\begin{align*}
&c_1^2 I \preceq \hat P(k_0, x_{k_0}) \preceq c_2^2 I, \\
&\bE \left[ \frac{\partial^\top f_{k_0} (x_{k_0},\xi_{k_0})}{\partial x} \hat P(k_0+1, f_{k_0} (x_{k_0}, \xi_{k_0} ))\right. \\
&\hspace{5mm}\left. \frac{\partial f_{k_0} (x_{k_0}, \xi_{k_0} )}{\partial x } \right]
\preceq \lambda^2 \hat P(k_0, x_{k_0})
\end{align*}
for all~$(k_0, x_{k_0}) \in \bZ \times \bR^n$. 
\red
\end{seccor}

We further consider a stationary case, i.e., $\xi_k$ and $f_k$ are independent of~$k$.
\begin{secasm}\label{stationary:asm}
The stochastic process~$\xi$ is stationary (in the strict sense), i.e., none of the characteristics of~$\xi_k$ changes with time~$k$. Moreover, none of~$f_k$ changes with time~$k$, i.e., $\hat f_0 = f_k$ for all~$k \in \bZ$. 
\red
\end{secasm}

Note that the stochastic process satisfying Assumptions~\ref{TI:asm} and~\ref{stationary:asm} is an i.i.d. process. Under Assumptions~\ref{TI:asm} and~\ref{stationary:asm}, $\hat P$ in~\eqref{Phat} can be chosen as a $k_0$-independent function. Namely, we have the following corollary without the proof.

\begin{seccor}\label{IES_iid_Euclid:cor}
Suppose that Assumptions~\ref{TI:asm} and~\ref{stationary:asm} hold. A system~\eqref{sys} is UIES in the second moment with respect to the Euclidean distance if and only if there exist $c_1, c_2 > 0$, $\lambda \in (0,1)$, and $\hat P_0: \bR^n \to \bS_{\succ 0}^{n \times n}$ such that
\begin{align}
&c_1^2 I \preceq \hat P_0 (x_0) \preceq c_2^2 I, \nonumber\\
&\bE \left[ \frac{\partial^\top \hat f_0 (x_0, \xi_0 )}{\partial x} \hat P_0( \hat f_0 (x_0, \xi_0)) \frac{\partial \hat f_0 (x_0, \xi_0 )}{\partial x} \right] 
\preceq \lambda^2 \hat P_0 (x_0) \label{IES_iid_Euclid:cond}
\end{align}
for all~$x_0 \in \bR^n$. 
\red
\end{seccor}

\begin{secrem}
In Corollary~\ref{IES_iid_Euclid:cor}, we have considered the stationary case. As a more general case, Corollary~\ref{IES_TI_Euclid:cor} can be specialized to the periodic case where there exists a positive integer~$N$ such that $f_{\kappa N+i} = f_{\kappa+i}$, $i=0, 1, \dots, N-1$, $\kappa \in \bZ$ and none of the characteristics of~$\xi_{\kappa N+i}$, $i=0,1,\dots, N-1$ changes with~$\kappa \in \bZ$. The generalized condition is described by using periodic~$\hat P_i$, $i=0, 1,\dots, N-1$; for more details, see a similar discussion in the linear case \cite[Corollary 3]{HH:22}.
\red
\end{secrem}

\subsection{General Markov Processes}
In this subsection, we consider the case where~$\xi$ is a general \emph{Markov process}.
\begin{secasm}\label{Markov:asm}
For each~$\Theta_j \subset \bR^m$, every~$j \in \bZ_{(i+1)+}$ and~$i \in \bZ$, it follows that
\begin{align*}
\bP (\xi_j \in \Theta_j | \xi_i, \xi_{i-1}, \dots ) = \bP (\xi_j \in \Theta_j  | \xi_i),
\end{align*}
where~$\bP (\cdot | \cdot )$ denotes the conditional probability.
\red
\end{secasm}

Assumption~\ref{Markov:asm} implies that the conditional expectation~$\bE_0$ can be simplified as
\begin{align*}
\bE_0[ \cdot ]  &= \bE [\cdot | \xi_{k_0-1} = \hat \xi_{k_0-1} ],\\
\bE_0 [ \cdot | \cF_k] &= \bE [\cdot | \xi_k], \; k \in \bZ_{k_0+}
\end{align*}
for each~$(k_0, \hat \xi_{k_0-1} ) \in \bZ \times \Theta_{k_0-1}$, where note that~$\Theta_{k_0-1}$ is the support of~$\xi_{k_0-1}$.  Then, for $P$ in Theorem~\ref{IES_Euclid:thm}, there exists~$\hat P:\bZ \times \bR^n \times \bR^m \to \bS_{\succ 0}^{n \times n}$ such that
\begin{align*}
&\bE_0 [P(k_0, x_{k_0}, S_{k_0} \xi^{k_0+})]\\
&=\bE [P(k_0, x_{k_0}, S_{k_0} \xi^{k_0+}) | \xi_{k_0-1} = \hat \xi_{k_0-1}  ]\\
&=\hat P(k_0, x_{k_0}, \hat \xi_{k_0-1} )
\end{align*}
for each~$(k_0, x_{k_0}, \hat \xi_{k_0-1} ) \in \bZ \times \bR^n \times \Theta_{k_0-1}$. Therefore, we have the following corollary of Theorem~\ref{IES_Euclid:thm} for Markov processes without the proof. 

\begin{seccor}\label{IES_Markov_Euclid:cor}
Suppose that Assumption~\ref{Markov:asm} holds.
A system~\eqref{sys} is UIES in the second moment with respect to the Euclidean distance if and only if there exist $c_1, c_2 > 0$, $\lambda \in (0,1)$, and $\hat P: \bZ \times \bR^n \times \bR^m \to \bS_{\succ 0}^{n \times n}$ such that
\begin{align*}
&c_1^2 I \preceq \hat P(k_0, x_{k_0}, \hat \xi_{k_0-1} ) \preceq c_2^2 I, \\
&\bE \left[ \frac{\partial^\top f_{k_0} (x_{k_0}, \xi_{k_0})}{\partial x} \right. 
\hat P(k_0+1, f_{k_0} (x_{k_0}, \xi_{k_0}), \xi_{k_0}) \\
&\hspace{5mm}\left. \frac{\partial f_{k_0} (x_{k_0}, \xi_{k_0})}{\partial x} \biggl| \xi_{k_0-1} =\hat \xi_{k_0-1} \right]
\preceq \lambda^2 \hat P(k_0, x_{k_0}, \hat \xi_{k_0-1} )
\end{align*}
for all~$(k_0, x_{k_0}, \hat \xi_{k_0-1} ) \in \bZ \times \bR^n \times \Theta_{k_0-1}$. 
\red
\end{seccor}

In the stationary case, again $\hat P$ can be chosen as a $k_0$-independent function. Namely, we have the following corollary without the proof.
\begin{seccor}\label{IES_Markov_stationary_Euclid:cor}
Suppose that Assumptions~\ref{stationary:asm} and~\ref{Markov:asm} hold. A system~\eqref{sys} is UIES in the second moment with respect to the Euclidean distance if and only if there exist $c_1, c_2 > 0$, $\lambda \in (0,1)$, and $\hat P_0: \bR^n \times \bR^m \to \bS_{\succ 0}^{n \times n}$ such that
\begin{align*}
&c_1^2 I \preceq \hat P_0 (x_0, \hat \xi_{-1} ) \preceq c_2^2 I, \\
&\bE \left[ \frac{\partial^\top \hat f_0 (x_0, \xi_0)}{\partial x} 
\hat P_0 (\hat f_0 (x_0, \xi_0), \xi_0) 
\frac{\partial \hat f_0 (x_0, \xi_0)}{\partial x} \biggl| \xi_{-1} =\hat \xi_{-1} \right]\\
&\preceq \lambda^2 \hat P_0 (x_0, \hat \xi_{-1} )
\end{align*}
for all~$(x_0, \hat \xi_{-1} ) \in \bR^n \times \Theta_{-1}$. 
\red
\end{seccor}

\begin{secrem}
Again Corollary~\ref{IES_Markov_Euclid:cor} can be specialized to the periodic case by using periodic~$\hat P_i$, $i=0, 1,\dots, N-1$.
\red
\end{secrem}

\subsection{Finite-mode Markov Chains}
In this subsection, we further consider the case where $\xi$ is a finite-mode Markov chain, which is non-stationary (i.e., non-homogeneous) unless the transition probability is time-invariant.
\begin{secasm}\label{finite_Markov:asm}
The process $\xi$ is given by a finite-mode Markov chain defined on the mode set $\cM :=\{1,...,M\}$, i.e., $\xi_k$ can take a value only in $\cM$ at each $k\in\bZ$.
\red
\end{secasm}

The process $\xi$ satisfying this assumption is a special case of the general Markov process in Assumption~\ref{Markov:asm}, and the corresponding system \eqref{sys} can be seen as a stochastic switched nonlinear system with the switching signal $s=\xi$ in Remark~\ref{switch:rem} given by a finite-mode Markov chain. Such a system is nothing but a standard Markov jump nonlinear system, e.g.,~\cite{GC:13,ZHZ:14}; also see, e.g.,~\cite{lLD:06} for Markov jump linear systems. This exemplifies the generality of the system class dealt with in this paper.

Let us denote the transition probability from mode $i$ to $j$ by
\begin{align*}
\pi_{j,i}^{k}:= \bP (\xi_{k+1} = j  | \xi_{k} = i) \ge 0
\end{align*}
for each $k\in\bZ$. By the definition, it satisfies
\begin{align*}
\sum_{j \in \cM} \pi_{j,i}^{k} = 1, \; k \in \bZ
\end{align*}
for all $i \in \cM$. Then, by using the mode dependent function $\hat P_i$,
$i \in \cM$,
Corollaries~\ref{IES_Markov_Euclid:cor} and \ref{IES_Markov_stationary_Euclid:cor} are further simplified as stated below, where the latter is about the stationary (i.e., homogeneous) Markov chain.

\begin{seccor}\label{IES_finite_Markov_Euclid:cor}
Suppose that Assumption~\ref{finite_Markov:asm} holds.
A system~\eqref{sys} is UIES in the second moment with respect to the Euclidean distance if and only if there exist $c_1, c_2 > 0$, $\lambda \in (0,1)$, and $\hat P_i: \bZ \times \bR^n \to \bS_{\succ 0}^{n \times n}$, $i \in \cM$ such that
\begin{align*}
&c_1^2 I \preceq \hat P_i(k_0, x_{k_0}) \preceq c_2^2 I, \\
&\sum_{j \in \cM}  \pi_{j,i}^{k_0} \frac{\partial^\top f_{k_0} (x_{k_0}, j)}{\partial x} 
\hat P_j(k_0+1, f_{k_0} (x_{k_0}, j) ) \\
&\hspace{13mm} \frac{\partial f_{k_0} (x_{k_0}, j)}{\partial x} 
\preceq \lambda^2 \hat P_i (k_0, x_{k_0} )
\end{align*}
for all~$(k_0, x_{k_0}, i) \in \bZ \times \bR^n \times \cM$.
\red
\end{seccor}

\begin{seccor}\label{IES_Markov_Jump_Euclid:cor}
Suppose that Assumptions~\ref{stationary:asm} and~\ref{finite_Markov:asm} hold. 
A system~\eqref{sys} is UIES in the second moment with respect to the Euclidean distance if and only if there exist $c_1, c_2 > 0$, $\lambda \in (0,1)$, and $\hat P_{0,i}: \bR^n \to \bS_{\succ 0}^{n \times n}$, $i\in \cM$ such that
\begin{align}
&c_1^2 I \preceq \hat P_{0,i} (x_0 ) \preceq c_2^2 I, \nonumber\\
&\sum_{j \in \cM}  \pi_{j,i} \frac{\partial^\top \hat f_0 (x_0, j)}{\partial x} \hat P_{0,j} (\hat f_0 (x_0, j))  \frac{\partial \hat f_0 (x_0, j)}{\partial x} \preceq \lambda^2 \hat P_{0,i} (x_0 ) \label{Markovjump_IES:cond}
\end{align}
for all~$(x_0, i) \in \bR^n \times \cM$. 
\red
\end{seccor}

\begin{secrem}
Again Corollary~\ref{IES_finite_Markov_Euclid:cor} can be specialized to the periodic case by using periodic~$\hat P_{k,i}$, $k=0, 1,\dots, N-1$.
\red
\end{secrem}

In the linear case, i.e., $\hat f_0 (x, j) = A_j x$,  the inequality~\eqref{Markovjump_IES:cond} reduces to an inequality that is equivalent to \cite[Equation (3.15)]{CFM:05}. By restricting classes of the processes and systems, we finally establish the connection between our results and the well known condition for Markov jump linear systems.

\section{Examples}\label{Ex:sec}
\subsection{Stabilizing Controller Design for Mechanical Systems}
In this subsection, the proposed stability condition for an i.i.d. process is applied to stabilizing controller design. Consider a pendulum controlled by a DC motor:
\begin{align*}
\left\{\begin{array}{l}
\ddot x_p + d \dot x_p + \xi_0 \sin(x_p) = a x_i,\\
L \dot x_i + R x_i + k_v \dot x_p = u,\\
\end{array}\right.
\end{align*}
where~$x_p$ and~$x_i$ denote the position of the mass and the current of the circuit, respectively. The control input~$u$ is the voltage. The parameter $\xi_0$ is unknown, and suppose that it follows the i.i.d. uniform distribution $U[1,2]$. The other parameters are~$d = 5$, $a = 2$, $k_v/L = 2$, $R/L = 5$, and~$1/L = 10$. We take the state as~$x := [x_p \; \dot x_p \; x_i]^\top$. Then, the Euler forward discretization of its state-space representation with the sampling period~$\Delta T = 1/20$ is
\begin{align*}
&x_{k+1} = A x_k + \hat f (x_k, \xi_0) + B  u_k,\\
&\hspace{4mm}A 
:=\begin{bmatrix}
1 &1/20 & 0 \\
0 & 3/4 & 1/10 \\
0 &-1/10 & 3/4
\end{bmatrix}, \;
B:= 
\begin{bmatrix}
0 \\ 0 \\ 1/2
\end{bmatrix}\\
&\hspace{4mm}\hat f(x,\xi_0) :=
\begin{bmatrix}
0\\ - \xi_0 \sin (x_p)/20 \\ 0
\end{bmatrix}, \; \frac{\partial \hat f (x_0, \xi_0)}{\partial x} :=  \xi_0 F(x),\\
&\hspace{4mm}
F(x):=
\begin{bmatrix}
0 & 0 & 0\\ - \cos (x_p)/20 & 0 & 0\\ 0 & 0 & 0
\end{bmatrix}. 
\end{align*}
This system satisfies Assumptions~\ref{TI:asm} and~\ref{stationary:asm}, and thus Corollary~\ref{IES_iid_Euclid:cor} is applicable. 

Let us fix $\hat P_0(x)$ on a constant. For stabilizing controller design, \eqref{IES_iid_Euclid:cond} becomes
\begin{align*}
\lambda^2 \hat P_0 &\succeq \bE [ ( A + \xi_0 F(x) + B K )^\top \hat P_0 ( A + \xi_0 F(x) + B K )] \\
&= ( A + \bE[\xi_0] F(x) + B K)^\top \hat P_0 ( A + \bE[\xi_0] F(x)  + B K) \\ 
&\hspace{5mm}+ (\bE[\xi_0^2] - (\bE [\xi_0])^2) F^\top (x) \hat P_0 F (x)
\end{align*}
where $K \in \bR^{1 \times 3}$ denotes a feedback gain. Utilizing the Schur complement technique with $\hat P_0 \succ 0$ and introducing new variables $\tilde K := K \hat P_0^{-1}$ and $\tilde P_0:=\hat P_0^{-1}$ yield the following equivalent LMI:
\begin{align*}
\begin{bmatrix}
\lambda^2 \tilde P_0  &*  & * \\
(A + \bE [\xi_0] F(x) ) \tilde P_0 + B \tilde K &  \tilde P_0 & * \\
\sqrt{\bE[\xi_0^2] - (\bE [\xi_0])^2} F(x) \tilde P_0 & 0 & \tilde P_0
\end{bmatrix} \succeq 0,
\end{align*}
where $*$ represents an appropriate matrix. This is an infinite family of linear matrix inequalities (LMIs). 

\begin{figure}[tb]
\begin{center}
\includegraphics[width=80mm]{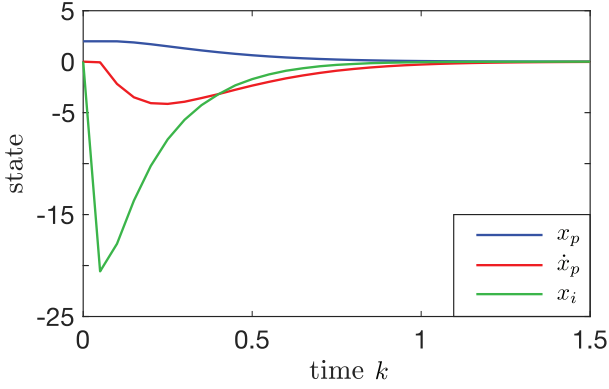}
\caption{State trajectory of the closed-loop system}
\label{mech:fig}
\end{center}
\end{figure}

It is well known that an infinite family of LMIs can be reduced to a finite one by a convex relaxation. Let us introduce
\begin{align*}
F^{(1)} :=  \begin{bmatrix}
0 &0 & 0 \\
-1/20 & 0 & 0 \\
0 & 0 & 0
\end{bmatrix} , \; 
F^{(2)} :=  \begin{bmatrix}
0 &0 & 0 \\
1/20 & 0 & 0 \\
0 & 0 & 0
\end{bmatrix}.
\end{align*}
Then, for each~$x_p \in \bR$, there exist $\theta^{(\ell )} (x_p) \in [0,1]$, $\ell =1,2$ such that
\begin{align*}
&\theta^{(1)} (x_p) + \theta^{(2)} (x_p) = 1,\\
&F(x_p) = \theta^{(1)} (x_p) F^{(1)} + \theta^{(2)} (x_p) F^{(2)}.
\end{align*}
Therefore, for stabilizing controller design, it suffices to solve the following set of LMIs:
\begin{align*}
\left\{\begin{array}{l}
 \tilde P_0 \succ 0,\\
\begin{bmatrix}
\lambda^2 \tilde P_0  &*  & * \\
(A + \bE [\xi_0] F^{(\ell)} ) \tilde P_0 + B \tilde K &  \tilde P_0 & * \\
\sqrt{\bE[\xi_0^2] - (\bE [\xi_0])^2} F^{(\ell)} \tilde P_0 & 0 & \tilde P_0
\end{bmatrix} \succeq 0, \; \ell =1, 2,
\end{array}\right. 
 \end{align*}
where $ \bE [\xi_0] =3/2$ and $\sqrt{\bE[\xi_0^2] - (\bE[\xi_0])^2} = 1/\sqrt{12}$ for the uniform distribution $U[1,2]$. Solving this for~$\lambda = \sqrt{0.9}$ gives a stabilizing feedback gain:
\begin{align*}
K = \begin{bmatrix} -20.6  & -4.09 &  -1.75 \end{bmatrix}.
\end{align*}
Fig.~\ref{mech:fig} shows a sample trajectory of the closed-loop system starting from $x(0) = [2 \; 0 \; 0]^\top$. It is confirmed that the closed-loop system is stabilized at the origin. In this example, we have assumed that the other parameters than~$\xi_0$ are deterministic. When they follow some probability distributions, controller design can be done similarly by generalizing the systematic methodology for linear systems~\cite{HPH:20}.

\subsection{Observer Design for Markov Jump Systems}
For the continuous-time deterministic system, $\dot x = f(x)$, $y=Cx$, it is known that if there exists a matrix $H$ making $\dot {\hat x} = f(\hat x) + H C\hat x$ IES, then $\dot {\hat x} = f(\hat x) + H (C\hat x - y)$ is an observer of the system; see, e.g.,~\cite{KBC:20}. This result can be extended to discrete-time stochastic systems, which is used for observer design of a Markov jump system.

Consider the following Markov jump system:
\begin{align*}
\left\{\begin{array}{l}
x_{k+1} = \hat f_0 (x_k,j ) :=
\begin{bmatrix}
x_{1,k} + x_{2,k} \\ \hat f_{0,2}(x_{1,k}, j) 
\end{bmatrix}\\
y_k = C x_k, \; 
C=\begin{bmatrix}
1 & 1 
\end{bmatrix}\\
\bP (\xi_{k+1} = i  | \xi_{k} = j) =\pi_{i,j},
\end{array}\right.
\end{align*}
where $i, j = 1, 2, 3$ (i.e., three modes) and
\begin{align*}
&\frac{d\hat f_{0,2}(z, 1)}{dz} \in [-3/4, -1/4], \; 
\frac{d\hat f_{0,2}(z, 2)}{dz} \in [-1/4, 1/4], \\
&\frac{d\hat f_{0,2}(z, 3)}{dz} \in [1/4, 3/4], \;
(\pi_{i,j})_{i,j}=\begin{bmatrix}
0.1 & 0.5 & 0.3\\
0.8 & 0.5 & 0.1\\
0.1 & 0   & 0.6
\end{bmatrix}.
\end{align*}
This system satisfies Assumptions~\ref{stationary:asm} and~\ref{finite_Markov:asm}, and thus we can utilize Corollary~\ref{IES_Markov_Jump_Euclid:cor} for observer design. 

Let us fix $\hat P_{0,j} (x_0)$, $j=1,2,3$ on constants. For observer design, \eqref{Markovjump_IES:cond} becomes
\begin{align*}
&\sum_{j \in \cM}  \pi_{j,i} \left(\frac{\partial \hat f_0 (x_0, j)}{\partial x} +H_j C \right)^\top \hat P_{0,j}\\
&\hspace{13mm} \left( \frac{\partial \hat f_0 (x_0, j)}{\partial x} + H_j C \right) \preceq \lambda^2 \hat P_{0,i}, \;  i = 1, 2, 3,
\end{align*}
where~$H_j \in \bR^3$ denotes an observer gain at each mode. Utilizing the Schur complement technique with $\hat P_{0,j} \succ 0$, $j=1,2,3$ and introducing new variables $\hat H_j := \hat P_{0,j} H_j$, $j=1,2,3$ yield the following equivalent LMI:
\begin{align*}
&\begin{bmatrix}
\lambda^2 \hat P_{0,i} & * & *  & *  \\
\displaystyle \sqrt{\pi_{1,i}} \left( \hat P_{0,1} \frac{\partial \hat f_0 (x_0, 1)}{\partial x} + \hat H_1 C \right) & \hat P_{0,1} & * & * \\[2mm]
\displaystyle \sqrt{\pi_{2,i}} \left( \hat P_{0,2} \frac{\partial \hat f_0 (x_0, 2)}{\partial x} + \hat H_2 C \right) & 0 & \hat P_{0,2}  & * \\[2mm]
\displaystyle \sqrt{\pi_{3,i}} \left(\hat P_{0,3} \frac{\partial \hat f_0 (x_0, 3)}{\partial x} + \hat H_3 C \right) & 0 & 0 & \hat P_{0,3}
\end{bmatrix}
\succeq 0,
\end{align*}
where $*$ represents an appropriate matrix. This is an infinite family of LIMs, which can be reduced to a finite one through a convex relaxation. Now, we define
\begin{align*}
A_1^{(1)} &:= 
\begin{bmatrix}
1 & 1 \\  -3/4 & 0
\end{bmatrix}, \;
A_1^{(2)} := 
A_2^{(1)} := 
\begin{bmatrix}
1 & 1 \\  -1/4 & 0
\end{bmatrix}, \\
A_2^{(2)} &:= 
A_3^{(1)} := 
\begin{bmatrix}
1 & 1 \\  1/4 & 0
\end{bmatrix}, \;
A_3^{(2)} := 
\begin{bmatrix}
1 & 1 \\  3/4 & 0
\end{bmatrix}.
\end{align*}
Then for each~$x_1 \in \bR$ and every~$j = 1,2,3$, there exist~$\theta^{(\ell )}(x_1, j) \in [0, 1]$, $\ell =1,2$ such that
\begin{align*}
&\theta^{(1)}(x_1,j)+\theta^{(2)}(x_1,j) = 1,\\
&\frac{\partial \hat f_0 (x_1, j)}{\partial x} = \theta^{(1)}(x_1, j) A_j^{(1)} + \theta^{(2)}(x_1, j) A_j^{(2)}.
\end{align*}
Therefore, for observer design, it suffices to solve the following finite family of LMIs:
\begin{align}
&\left\{\begin{array}{l}
 \hat P_{0,i} \succ 0,\\
 \begin{bmatrix}
\lambda^2 \hat P_{0,i} & * & *   & * \\
\displaystyle \sqrt{\pi_{1,i}} \left( \hat P_{0,1} A_1^{(\ell)} + \hat H_1 C \right) & \hat P_{0,1} & *  & *  \\[2mm]
\displaystyle \sqrt{\pi_{2,i}} \left( \hat P_{0,2} A_2^{(\ell)} + \hat H_2 C \right) & 0 & \hat  P_{0,2}  & *  \\[2mm]
\displaystyle \sqrt{\pi_{3,i}} \left(\hat P_{0,3} A_3^{(\ell)} + \hat H_3 C \right) & 0 & 0 & \hat P_{0,3} 
\end{bmatrix}
\succeq 0,
\end{array}\right. \nonumber\\
&\hspace{60mm}i =1, 2, 3.\label{Markov:LMI}
\end{align}
By solving~\eqref{Markov:LMI} for~$\lambda = \sqrt{0.9}$, an observer gain at each mode is designed as follows:
\begin{align*}
H_1 &:= \hat P_{0,1}^{-1} \hat H_1 
= 
\begin{bmatrix}
-1.00 \\ 0.824
\end{bmatrix}, \; 
H_2 := \hat P_{0,2}^{-1} \hat H_2 
= 
\begin{bmatrix}
-1.00 \\ 0.338
\end{bmatrix}, \\
H_3 &:= \hat P_{0,3}^{-1} \hat H_3
= 
\begin{bmatrix}
-1.00 \\ 0.00470
\end{bmatrix}.
\end{align*}

Next, we consider finding a common observer gain $\hat H$ for all modes using a common $\hat P_0$. This can be done simply by substituting $\hat P_{0,j} = \hat P_0$ and $\hat H_j = \hat H$, $j=1,2,3$ into~\eqref{Markov:LMI}. Solving the corresponding set of LMIs for~$\lambda = \sqrt{0.9}$, a common observer gain is designed as follows:
\begin{align*}
H := \hat P_0^{-1} \hat H 
= 
\begin{bmatrix}
0.0751 \\ -0.00214
\end{bmatrix}.
\end{align*}

\begin{figure}[tb]
\begin{center}
\includegraphics[width=80mm]{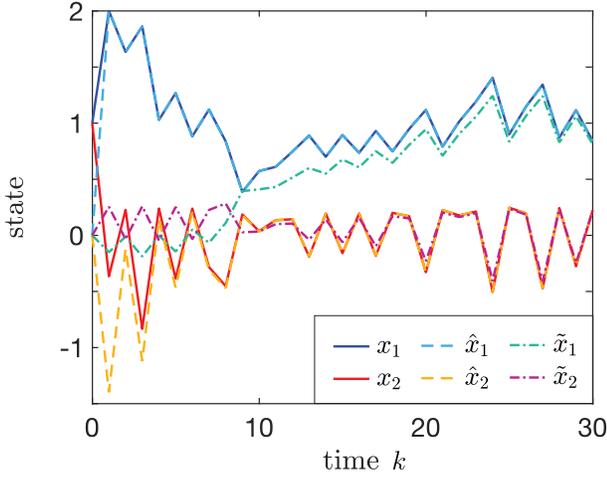}
\caption{Trajectories of the system and its observers}
\label{Markov:fig}
\end{center}
\end{figure}

For the sake of simulation, we choose
\begin{align*}
\hat f_{0,2}(z, 1) &= \cos (z)/4 - z/2, \;
\hat f_{0,2}(z, 2) = \sin (z)/4, \\
\hat f_{0,2}(z, 3) &= \cos (z)/4 + z/2.
\end{align*}
Fig.~\ref{Markov:fig} shows a sample trajectory of the system starting from $x(0) = [2 \; 2]^\top$ and the trajectories of its mode-dependent observer starting from $\hat x(0) = [0 \; 0]^\top$ and its mode-independent observer starting from $\tilde x(0) = [0 \; 0]^\top$. It is confirmed that the trajectories of both observers converge to that of the system, and faster convergence is achieved by the mode-dependent observer. 

\section{Conclusion}\label{Con:sec}
In this paper, we have studied moment UIES for discrete-time nonlinear
stochastic systems in the contraction framework. In particular, we have
presented a sufficient condition for UIES in the first moment with
respect to the Riemannian metric and a necessary and sufficient condition for UIES in the second moment with respect to the Euclidean distance. Then, the second moment UIES condition has been applied to i.i.d. processes and Markov processes as specialized applications. Future work includes developing general control/observer design methods, partly illustrated by this paper, in the proposed contraction framework.

\appendices
\section{Proof of Theorem~\ref{IES_dtm:thm}}\label{IES_dtm:app}
\begin{IEEEproof}
(Step 1)
From~\eqref{sys_dtm_sol} and~\eqref{IES_dtm:cond}, the variational system~\eqref{vsys_dtm} satisfies the following:
\begin{align*}
&\delta z_{k+1}^\top P(k+1, \psi_{k+1}) \delta z_{k+1}\\
&= \delta z_k^\top \frac{\partial^\top g_k (\psi_k)}{\partial \psi_k} P(k+1,g_k ( \psi_k)) \frac{\partial g_k (\psi_k)}{\partial \psi_k} \delta z_k\\
&\le \lambda^2 \delta z_k^\top P(k, \psi_k) \delta z_k, \; \forall k \in \bZ_{k_0+},
\end{align*}
where the arguments of~$\psi_k(k_0, z_{k_0})$ are omitted.
Repeating this leads to
\begin{align*}
\delta z_k^\top P(k, \psi_k(k_0, z_{k_0})) \delta z_k \le \lambda^{2(k-k_0)} \delta z_{k_0}^\top P(k_0, z_{k_0}) \delta z_{k_0}.
\end{align*}
Furthermore, applying~\eqref{Pbd_dtm:cond2} to both sides and taking the square roots of them yield
\begin{align}
&\sqrt{\delta z_k^\top \hat P( \psi_k(k_0, z_{k_0})) \delta z_k} \nonumber\\
& \le \frac{c_2}{c_1} \lambda^{k-k_0} \sqrt{ \delta z_{k_0}^\top \hat P( z_{k_0}) \delta z_{k_0}}, \; \forall k \in \bZ_{k_0+}
\label{IES_dtm:pf1}
\end{align}
for each~$(k_0, z_{k_0}) \in \bZ \times \bR^n$.

(Step 2)
In this step, we introduce the considered distance function.
For any pair~$(z', z'') \in \bR^n \times \bR^n$, let $\Gamma (z', z'')$ denote the collection of piecewise $C^1$ paths~$\gamma:[0,1] \to \bR^n$ such that~$\gamma (0) =z'$ and~$\gamma (1)=z''$. Then, we define a non-negative function~$d_{\hat P} :\bR^n \times \bR^n \to \bR$ by
\begin{align}
d_{\hat P} (z', z'') := \inf_{\gamma \in \Gamma (z', z'')} \int_0^1 \sqrt{\frac{d^\top \gamma (s)}{ds} \hat P(\gamma (s)) \frac{d \gamma (s)}{ds}} ds,
\label{distance:def}
\end{align}
where~$d \gamma (s)/ds$ stands for a semi-differentiation.
This is nothing but a distance function. The Hopf-Rinow theorem~\cite[Theorem 6.6.1]{BCS:12} guarantees that for each~$(z', z'') \in \bR^n \times \bR^n$, there exists a geodesic~$\gamma^* \in \Gamma(z', z'')$, i.e.,
\begin{align}
d_{\hat P} (z', z'') =  \int_0^1 \sqrt{ \frac{d^\top \gamma^* (s)}{ds} \hat P(\gamma^* (s)) \frac{d \gamma^* (s)}{ds}} ds. 
\label{distance_upbd}
\end{align}

(Step 3)
We relate the geodesic~$\gamma^* \in \Gamma (z'_{k_0},z''_{k_0})$ to the variational system~\eqref{vsys_dtm}. Let us choose the initial states of the system~\eqref{sys_dtm} and variational system~\eqref{vsys_dtm} at the initial time~$k_0 \in \bZ$ as~$(z_{k_0},\delta z_{k_0})=(\gamma^* (s), d\gamma^* (s)/ds)$, $s \in [0,1]$; note that~$\gamma^*$ is independent of~$k_0$ and depends only on~$(z'_{k_0},z''_{k_0})$. Then, it follows that
\begin{align}
\frac{\partial \psi_{k+1}(k_0, \gamma^* (s))}{\partial s} \nonumber
&=\frac{\partial g_k(\psi_k (k_0, \gamma^* (s)))}{\partial s} \nonumber\\
&=\frac{\partial g_k (\psi_k )}{\partial \psi_k} \frac{\partial \psi_k(k_0, \gamma^* (s))}{\partial s}, \; \forall k \in \bZ_{k_0+}
\label{IES_dtm:pf2}
\end{align}
for each~$(k_0, (z'_{k_0},z''_{k_0})) \in \bZ \times (\bR^n \times \bR^n)$ and every~$s \in [0,1]$, where the first equality is obtained from~\eqref{sys_dtm_sol}; the second one is by the chain rule. This implies that~$\partial \psi_k(k_0, \gamma^* (s))/\partial s$ satisfies the dynamics of the variational system~\eqref{vsys_dtm} and consequently~\eqref{IES_dtm:pf1}. That is, we have
\begin{align*}
&\sqrt{\frac{\partial^\top \psi_k(k_0, \gamma^* (s))}{\partial s} \hat P(\psi_k(k_0, \gamma^*(s))) \frac{\partial \psi_k(k_0, \gamma^* (s))}{\partial s}}\\
& \le  \frac{c_2}{c_1} \lambda^{k-k_0} \sqrt{\frac{d^\top \gamma^* (s)}{ds} \hat P(\gamma^*(s)) \frac{d\gamma^* (s)}{ds}}, \; \forall k \in \bZ_{k_0+}
\end{align*}
for each~$(k_0, (z'_{k_0},z''_{k_0})) \in \bZ \times (\bR^n \times \bR^n)$ and every~$s \in [0,1]$.
Note that~$\psi_k(k_0, \gamma^* (s))$, $k \in \bZ_{k_0+}$ is a path connecting $\psi_k(k_0, \gamma^* (0))=\psi_k(k_0, z'_{k_0})$ to~$\psi_k(k_0, \gamma^* (1))=\psi_k(k_0, z''_{k_0})$. Therefore, from~\eqref{distance:def} and~\eqref{distance_upbd}, integrating both sides with respect to~$s$ in the interval~$[0,1]$ leads to
\begin{align*}
&d_{\hat P}(\psi_k(k_0, z'_{k_0}), \psi_k(k_0, z''_{k_0}))\\
&\le  \frac{c_2}{c_1} \lambda^{k-k_0} d_{\hat P}(z'_{k_0}, z''_{k_0}), \; \forall k \in \bZ_{k_0+}
\end{align*}
for each~$(k_0, (z'_{k_0}, z''_{k_0})) \in \bZ \times (\bR^n \times \bR^n)$.
\end{IEEEproof}
\section{Proof of Theorem~\ref{IES:thm}}\label{IES:app}
Before providing the proof, we proceed with auxiliary analysis for the variational system~\eqref{vsys}. Its solution can be described as
\begin{align}
\delta x_k = \Phi_k (\xi^{(k-1)-}; k_0, x_{k_0}, \hat \xi^{(k_0-1)-}) \delta x_{k_0}, \; k \in \bZ_{k_0+}, \label{Phi1}
\end{align}
or simply $\delta x_k = \Phi_k (\xi^{(k-1)-}) \delta x_{k_0}$, $k \in \bZ_{k_0+}$ for each $(k_0, (x_{k_0}, \delta x_{k_0}), \hat \xi^{(k_0-1)-}) \in \bZ \times (\bR^n  \times \bR^n)  \times \hat \Xi^{(k_0-1)-}$, where using the solution~$\phi_k(\xi^{(k-1)-})$ to the system~\eqref{sys}, $\Phi_k$ is defined by
\begin{align}
&\Phi_{k_0} := \frac{\partial \phi_{k_0}}{\partial x_{k_0}}=I, \label{Phi2}\\
&\Phi_k ( \xi^{(k-1)-}; k_0, x_{k_0}, \hat \xi^{(k_0-1)-}) \nonumber\\
&: = \frac{\partial f_{k-1} (\phi_{k-1}(\xi^{(k-2)-}) , \xi_{k-1})}{\partial \phi_{k-1}} \cdots \nonumber\\
&\hspace{6mm} \frac{\partial f_{k_0+1} (\phi_{k_0+1}(\xi_{k_0}), \xi_{k_0+1})}{\partial \phi_{k_0+1}} \frac{\partial f_{k_0} (\phi_{k_0}, \xi_{k_0})}{\partial \phi_{k_0}} \frac{\partial \phi_{k_0}}{\partial x_{k_0}}, \nonumber\\ 
&\;= \frac{\partial  \phi_k (\xi^{(k-1)-}; k_0, x_{k_0}, \hat \xi^{(k_0-1)-})}{\partial x_{k_0}}, \; k \in \bZ_{(k_0+1)+}.
\label{Phi3}
\end{align}
In~\eqref{Phi2} or~\eqref{Phi3}, $\phi_{k_0} = x_{k_0}$, \eqref{sys_sol}, and the chain rule are used.

\begin{secrem}\label{mapping:rem}
The notations $\phi_k(\xi^{(k-1)-}; k_0,x_{k_0},\hat \xi^{(k_0-1)-})$ and $\Phi_k (\xi^{(k-1)-}; k_0, x_{k_0}, \hat \xi^{(k_0-1)-})$ (or $\phi_k(\xi^{(k-1)-})$ and $\Phi_k (\xi^{(k-1)-})$ in shorthand) are used to emphasize that they are stochastic processes. When they are considered as mappings $\phi_k: \bR^n \times (\bR^m)^{\bZ_{[k_0, k-1]}} \to \bR^n$ and $\Phi_k: \bR^n \times (\bR^m)^{\bZ_{[k_0, k-1]}} \to \bR^{n \times n}$, the notations~$\phi_k(x_{k_0}, \eta^{(k-1)-}; k_0)$ and~$\Phi_k(x_{k_0}, \eta^{(k-1)-}; k_0)$ with~$\eta$  (or $\phi_k$ and $\Phi_k$ in shorthand)  are used, respectively. Note that as the mappings, $\phi_k$ and $\Phi_k$ satisfy the counterparts of \eqref{sys_sol}, \eqref{Phi1}, and~\eqref{Phi3}, i.e.,
\begin{align}
&\phi_{k+1}(x_{k_0}, \eta^{k-}; k_0) = f_k(\phi_k(x_{k_0}, \eta^{(k-1)-}; k_0), \eta_k), \label{sys_sol_eta}\\
&\delta x_k = \Phi_k (x_{k_0}, \eta^{(k-1)-}; k_0) \delta x_{k_0}, \; k \in \bZ_{k_0+} \label{Phi1_eta}\\
&\Phi_k ( x_{k_0}, \eta^{(k-1)-}; k_0) = \frac{\partial  \phi_k (x_{k_0}, \eta^{(k-1)-}; k_0)}{\partial x_{k_0}},\label{Phi3_eta}\\
&\hspace{55mm} k \in \bZ_{(k_0+1)+}, \nonumber
\end{align}
 for each~$(k_0, (x_{k_0}, \delta x_{k_0}), \eta^{k-}) \in \bZ \times (\bR^n \times \bR^n) \times (\bR^m)^{\bZ_{[k_0, k]}}$, respectively.
 \red
\end{secrem}

\begin{secrem}\label{piecewise:rem}
Note that $\phi_k$, $k \in \bZ_{k_0+}$ is a composition function of~$f_i$, $i=k_0,\dots,{k-1}$. Under Assumption~\ref{f:asm}, both $\phi_k(x_{k_0}, \eta^{(k-1)-}; k_0)$, $k \in \bZ_{k_0+}$ and $\Phi_k(x_{k_0}, \eta^{(k-1)-}; k_0)$, $k \in \bZ_{k_0+}$ are piecewise continuous functions of $(x_{k_0}, \eta^{(k-1)-}) \in \bR^n \times (\bR^m)^{\bZ_{[k_0, k-1]}}$ at each~$k_0 \in \bZ$. Therefore, $\phi_k (\xi^{(k-1)-})$ and $\delta x_k = \Phi_k (\xi^{(k-1)-}) \delta x_{k_0}$ are both $\cF_{k-1}$-measurable functions for each $(k_0, (x_{k_0}, \delta x_{k_0}), \hat \xi^{(k_0-1)-}) \in \bZ \times (\bR^n \times \bR^n) \times \hat \Xi^{(k_0-1)-}$. 
\red
\end{secrem}

Now, we are ready to prove Theorem~\ref{IES:thm}.
\begin{IEEEproof}
(Step 1)
The quadratic forms of both sides in~\eqref{Pbd:cond} with respect to (the deterministic) $\delta x_{k_0} \in \bR^n$ satisfy
\begin{align}
c_1^2 \delta x_{k_0}^\top \hat P(x_{k_0}) \delta x_{k_0} &\le  \delta x_{k_0}^\top \bE_0 [P(k_0, x_{k_0}, S_{k_0} \xi^{k_0+}) ] \delta x_{k_0} \nonumber\\
&\le c_2^2  \delta x_{k_0}^\top \hat P(x_{k_0}) \delta x_{k_0}
\label{IES:pf2}
\end{align}
for each~$(k_0, (x_{k_0},\delta x_{k_0}), \hat \xi^{(k_0-1)-} ) \in \bZ \times (\bR^n \times \bR^n) \times \hat \Xi^{(k_0-1)-}$.  Since $(k_0, (x_{k_0},\delta x_{k_0}), \hat \xi^{(k_0-1)-} )$ is arbitrary in \eqref{IES:pf2}, and both $\phi_k(\xi^{(k-1)-})$ and $\delta x_k = \Phi_k (\xi^{(k-1)-}) \delta x_{k_0}$ are $\cF_{k-1}$-measurable as mentioned in Remark~\ref{piecewise:rem}, the inequality \eqref{IES:pf2} is preserved under time-shift: $k_0 \mapsto k$, $k \in \bZ_{k_0+}$. The time-shift: $k_0 \mapsto k$, $k \in \bZ_{k_0+}$ of the first inequality yields
\begin{align*}
&c_1^2 \delta x_k^\top \hat P(\phi_k (\xi^{(k-1)-})) \delta x_k \\
&\le  \delta x_k^\top \bE_0 [P(k, \phi_k (\xi^{(k-1)-}), S_k \xi^{k+}) | \cF_{k-1} ] \delta x_k  \mbox{ a.s.}, \\ 
&\hspace{60mm}\forall k \in \bZ_{k_0+}
\end{align*}
for each~$(k_0, (x_{k_0},\delta x_{k_0}), \hat \xi^{(k_0-1)-} ) \in \bZ \times (\bR^n \times \bR^n) \times \hat \Xi^{(k_0-1)-}$. 
Taking the conditional expectations~$\bE_0 [\cdot ]$ of both sides leads to
\begin{align}
&c_1^2 \bE_0 [\delta x_k^\top \hat P(\phi_k (\xi^{(k-1)-})) \delta x_k] \nonumber\\
&\le \bE_0 [ \delta x_k^\top \bE_0 [P(k, \phi_k (\xi^{(k-1)-}), S_k \xi^{k+}) | \cF_{k-1} ] \delta x_k ],\label{IES:pf3}\\
&\hspace{60mm}\forall k \in \bZ_{k_0+}\nonumber
\end{align}
for each~$(k_0, (x_{k_0},\delta x_{k_0}), \hat \xi^{(k_0-1)-} ) \in \bZ \times (\bR^n \times \bR^n) \times \hat \Xi^{(k_0-1)-}$. 

(Step~2) The quadratic forms of both sides in~\eqref{IES:cond} with respect to (the deterministic) $\delta x_{k_0} \in \bR^n$ satisfy 
\begin{align*}
&\bE_0 \left[ \delta x^\top_{k_0}  \frac{\partial^\top f_{k_0}(x_{k_0},\xi_{k_0})}{\partial x_{k_0}} \right. \\
&\hspace{7mm} \bE_0 [ P(k_0+1, f_{k_0}(x_{k_0}, \xi_{k_0}), S_{k_0+1} \xi^{(k_0+1)+})| \cF_{k_0}] \\
&\hspace{5mm}\left. \frac{\partial f_{k_0}(x_{k_0},\xi_{k_0})}{\partial x_{k_0}} \delta x_{k_0}\right] \\
&\le \lambda^2 \delta x^\top_{k_0} \bE_0[ P(k_0, x_{k_0}, S_{k_0} \xi^{k_0+}) ] \delta x_{k_0}
\end{align*}
for each~$(k_0, (x_{k_0},\delta x_{k_0}), \hat \xi^{(k_0-1)-} ) \in \bZ \times (\bR^n \times \bR^n) \times \hat \Xi^{(k_0-1)-}$. The time-shift: $k_0 \mapsto k$, $k \in \bZ_{k_0+}$ yields
\begin{align*}
&\bE_0 \left[ \delta x^\top_k  \frac{\partial^\top f_k (\phi_k (\xi^{(k-1)-}), \xi_k)}{\partial \phi_k} \right. \\
&\hspace{7mm} \bE_0 [ P(k+1, f_k (\phi_k (\xi^{(k-1)-}), \xi_k), S_{k+1} \xi^{(k+1)+})| \cF_k] \\
&\hspace{5mm}\left. \frac{\partial f_k (\phi_k (\xi^{(k-1)-}),\xi_k)}{\partial \phi_k} \delta x_k \Biggl| \cF_{k-1}\right] \\
&\le \lambda^2 \delta x^\top_k \bE_0[ P(k, \phi_k (\xi^{(k-1)-}), S_k \xi^{k+}) | \cF_{k-1}] \delta x_k \mbox{ a.s.},  \\
&\hspace{65mm}\forall k \in \bZ_{k_0+},
\end{align*}
or equivalently, from~\eqref{sys_sol} and~\eqref{vsys}, 
\begin{align*}
&\bE_0 [ \delta x^\top_{k+1} \bE_0 [ P(k+1, \phi_{k+1}(\xi^{k-}), S_{k+1} \xi^{(k+1)+})| \cF_k] \\
&\hspace{5mm}\delta x_{k+1} | \cF_{k-1} ] \\
&\le \lambda^2 \delta x^\top_k \bE_0[ P(k, \phi_k(\xi^{(k-1)-}), S_k \xi^{k+}) | \cF_{k-1}] \delta x_k \mbox{ a.s.}, \\
&\hspace{65mm} \forall k \in \bZ_{k_0+}
\end{align*}
for each~$(k_0, (x_{k_0},\delta x_{k_0}), \hat \xi^{(k_0-1)-} ) \in \bZ \times (\bR^n \times \bR^n) \times \hat \Xi^{(k_0-1)-}$. Recall that~$(\cF_k)_{k \in \bZ_{k_0+}}$ is a filtration on $(\Omega, \cF, \bP)$ for each~$\hat \xi^{(k_0-1)-} \in \hat \Xi^{(k_0-1)-}$. Then, taking the conditional expectations~$\bE_0[ \cdot ]$ of both sides of the above inequality yields
\begin{align*}
&\bE_0 [ \delta x^\top_{k+1} \bE_0 [ P(k+1, \phi_{k+1} (\xi^{k-}), S_{k+1} \xi^{(k+1)+})| \cF_k] \delta x_{k+1} ] \nonumber\\
&\le \lambda^2  \bE_0 [\delta x^\top_k\bE_0[ P(k, \phi_k (\xi^{(k-1)-}), S_k \xi^{k+}) | \cF_{k-1}] \delta x_k ], \nonumber\\
&\hspace{70mm} \forall k \in \bZ_{k_0+}
\end{align*}
for each~$(k_0, (x_{k_0},\delta x_{k_0}), \hat \xi^{(k_0-1)-} ) \in \bZ \times (\bR^n \times \bR^n) \times \hat \Xi^{(k_0-1)-}$. A recursive use of this leads to
\begin{align}
&\bE_0 [ \delta x^\top_k \bE_0 [ P(k, \phi_k (\xi^{(k-1)-}), S_k \xi^{k+})| \cF_{k-1}] \delta x_k ] \nonumber\\
&\le \lambda^{2(k-k_0)}  \delta x^\top_{k_0}  \bE_0[ P(k_0, x_{k_0} , S_{k_0}  \xi^{k_0 +}) ] \delta x_{k_0}, \; \forall k \in \bZ_{k_0+}
\label{IES:pf1}
\end{align}
for each~$(k_0, (x_{k_0},\delta x_{k_0}), \hat \xi^{(k_0-1)-} ) \in \bZ \times (\bR^n \times \bR^n) \times \hat \Xi^{(k_0-1)-}$. 

In summary, the second inequality of  \eqref{IES:pf2}, \eqref{IES:pf3}, and \eqref{IES:pf1} lead to
\begin{align}
&\bE_0 [ \delta x_k^\top \hat P(\phi_k (\xi^{(k-1)-})) \delta x_k ] \nonumber\\
& \le \frac{c_2^2}{c_1^2} \lambda^{2(k-k_0)} \delta x_{k_0}^\top \hat P(x_{k_0}) \delta x_{k_0}, \; \forall k \in \bZ_{k_0+}
\label{IES:pf4}
\end{align}
for each~$(k_0, (x_{k_0},\delta x_{k_0}), \hat \xi^{(k_0-1)-} ) \in \bZ \times (\bR^n \times \bR^n) \times \hat \Xi^{(k_0-1)-}$. 
Taking the square roots of both sides and applying the Cauchy--Schwarz inequality \cite[Corollary 3.1.12]{AL:06} (with~$\bP(\Omega) =1$) to the left-hand side yield
\begin{align}
&\bE_0 \left[ \sqrt{\delta x_k^\top \hat P(\phi_k (\xi^{(k-1)-})) \delta x_k} \; \right]\nonumber\\
&\le \frac{c_2}{c_1} \lambda^{k-k_0} \sqrt{\delta x_{k_0}^\top \hat P(x_{k_0}) \delta x_{k_0}}, \; \forall k \in \bZ_{k_0+}
\label{IES:pf5}
\end{align}
for each~$(k_0, (x_{k_0}, \delta x_{k_0}), \hat \xi^{(k_0-1)-}) \in \bZ \times (\bR^n \times \bR^n) \times \hat \Xi^{(k_0-1)-}$. 

(Step~3)
Here, we consider $\phi_k$ and $\Phi_k$ as the mappings $\phi_k(x_{k_0}, \eta^{(k-1)-}; k_0)$ and~$\Phi_k(x_{k_0}, \eta^{(k-1)-}; k_0)$; recall Remark~\ref{mapping:rem}. For each pair~$(x'_{k_0}, x''_{k_0}) \in \bR^n \times \bR^n$, let~$\gamma^* \in \Gamma (x'_{k_0}, x''_{k_0})$ be the geodesic with respect to~$\hat P$, i.e., a path satisfying~\eqref{distance_upbd}; note that $\gamma^*$ is independent of~$k_0$ and $\eta^{(k-1)-}$. As in~\eqref{IES_dtm:pf2}, let $(x_{k_0}, \delta x_{k_0}) = (\gamma^*(s), d\gamma^* (s)/ds)$, $s \in [0,1]$ be the initial states of $\phi_k(x_{k_0}, \eta^{(k-1)-}; k_0)$ and~$\Phi_k(x_{k_0}, \eta^{(k-1)-}; k_0)$. Then, it follows from \eqref{sys_sol_eta} and the chain rule that
\begin{align}
&\frac{\partial \phi_{k+1}( \gamma^* (s), \eta^{k-}; k_0)}{\partial s} \nonumber\\
&=\frac{\partial f_k(\phi_k(\gamma^* (s), \eta^{(k-1)-}; k_0), \eta_k)}{\partial s} \nonumber\\
&= \frac{\partial f_k(\phi_k (\gamma^* (s), \eta^{(k-1)-}; k_0), \eta_k)}{\partial \phi_k}
\frac{\partial \phi_k (\gamma^* (s), \eta^{(k-1)-}; k_0)}{\partial s}, \nonumber\\
&\hspace{60mm} \forall k \in \bZ_{k_0+} \label{vsys_path}
\end{align}
for each~$(k_0, (x'_{k_0}, x''_{k_0}), \eta^{k-}) \in \bZ \times (\bR^n \times \bR^n) \times ( \bR^m)^{\bZ_{[k_0,k]}}$ and every $s \in [0,1]$. This implies that $\partial \phi_k( \gamma^* (s), \eta^{(k-1)-}; k_0)/\partial s$ satisfies \eqref{Phi3_eta} for $(x_{k_0}, \delta x_{k_0}) = (\gamma^*(s), d\gamma^* (s)/ds)$, $s \in [0,1]$. In this case, \eqref{Phi1_eta} becomes
\begin{align}
&\frac{\partial \phi_k (\gamma^* (s), \eta^{(k-1)-}; k_0)}{\partial s} \nonumber\\
&=\Phi_k (\gamma^* (s), \eta^{(k-1)-}; k_0) \frac{d\gamma^* (s)}{ds}, \; \forall k \in \bZ_{k_0+}
\label{Phi_path_eta}
\end{align}
for each~$(k_0, (x'_{k_0}, x''_{k_0}), \eta^{(k-1)-}) \in \bZ \times (\bR^n \times \bR^n) \times ( \bR^m)^{\bZ_{[k_0,k-1]}}$ and every $s \in [0,1]$.

(Step~4)
Now, we consider stochastic processes. The equality~\eqref{vsys_path} implies that $\partial \phi_k (\xi^{(k-1)-}; k_0, \gamma^* (s), \hat \xi^{(k_0-1)-})/\partial s$, $k \in \bZ_{k_0+}$ is a solution to the variational system~\eqref{vsys} and satisfies the counterpart of~\eqref{Phi_path_eta}, i.e.,
\begin{align}
&\frac{\partial \phi_k (\xi^{(k-1)-}; k_0, \gamma^* (s), \hat \xi^{(k_0-1)-})}{\partial s} \nonumber\\
&=\Phi_k (\xi^{(k-1)-}; k_0, \gamma^* (s), \hat \xi^{(k_0-1)-}) \frac{d\gamma^* (s)}{ds}, \; \forall k \in \bZ_{k_0+}
\label{Phi_path}
\end{align}
under the initial state $(x_{k_0},\delta x_{k_0}) = (\gamma^*(s),d\gamma^*(s)/ds)$ for each~$(k_0, (x'_{k_0}, x''_{k_0}), \hat \xi^{(k_0-1)-}) \in \bZ \times (\bR^n \times \bR^n) \times \hat \Xi^{(k_0-1)-}$ and every $s \in [0,1]$.

Substituting $(x_{k_0}, \delta x_{k_0}) = (\gamma^*(s), d\gamma^* (s)/ds)$, $s \in [0,1]$ into~\eqref{IES:pf5} and applying~\eqref{Phi_path} lead to
\begin{align}
&\bE_0 \left[ \left( \frac{d^\top \gamma^*(s)}{ds} \Phi_k^\top (\xi^{(k-1)-}) \right. \right.\nonumber\\
&\hspace{10mm}\left. \left. \hat P(\phi_k(\xi^{(k-1)-})) \Phi_k (\xi^{(k-1)-}) \frac{d \gamma^*(s)}{ds} \right)^{1/2} \right]\nonumber\\
& \le \frac{c_2}{c_1} \lambda^{k-k_0} \sqrt{\frac{d^\top \gamma^*(s)}{ds} \hat P(\gamma^*(s)) \frac{d \gamma^*(s)}{ds}}, \; \forall k \in \bZ_{k_0+}
\label{IES:pf6}
\end{align}
for each~$(k_0, (x'_{k_0}, x''_{k_0}), \hat \xi^{(k_0-1)-}) \in \bZ \times (\bR^n \times \bR^n) \times \hat \Xi^{(k_0-1)-}$ and every $s \in [0,1]$, where in the left-hand side, the argument~$(k_0, \gamma^* (s), \hat \xi^{(k_0-1)-})$ is dropped from~$\phi_k (\xi^{(k-1)-})$ and $\Phi_k (\xi^{(k-1)-})$.

(Step 5)
We consider integrating both sides of~\eqref{IES:pf6} with respect to~$s$ in~$[0,1]$. In~\eqref{distance:def} and~\eqref{distance_upbd}, the Riemann integrals are used. For the sake of formality, they need to be replaced with the Lebesgue integrals. To this end, we introduce a measurable space corresponding to~$s$. Let~$(\bR, \cB(\bR), \mu)$ be the measurable space, where~$\mu$ is the Lebesgue measure. Note that both~$(\bR, \cB(\bR), \mu)$ and~$(\Omega, \cF, \bP)$ are complete and~$\sigma$-finite. Then, the product measurable space naturally induced by the Cartesian product~$\bR \times \Omega$, denoted by~$(\bR \times \Omega, \cL, \lambda)$, is complete and $\sigma$-finite~\cite[Theorem 5.1.2 and Remark 5.1.2]{AL:06}. 

To take the Lebesgue integrals for~\eqref{IES:pf6}, we introduce the following functions:
\begin{align}
( \overline{\gamma}(s) , \overline{\partial \gamma}(s) )
:= \left\{\begin{array}{cl}
( \gamma^* (s) , d \gamma^*(s)/ds ), & s \in [0,1]\\
(0,0) , & s \in \bR \setminus [0,1]
\end{array}\right. .
\label{new_gamma}
\end{align}
Since $\gamma^*$ is of piecewise $C^1$ on $[0,1]$, $( \overline{\gamma} , \overline{\partial \gamma} )$ is piecewise continuous on~$\bR$. In~\eqref{IES:pf6}, $( \gamma^* (s) , d \gamma^*(s)/ds )$ can be replaced with $( \overline{\gamma}(s) , \overline{\partial \gamma}(s) )$, and the corresponding inequality holds for all~$s \in \bR$ instead of~$s \in [0, 1]$. That is, we have
\begin{align}
&\bE_0 \left[ \left( \overline{\partial \gamma}^\top (s) \Phi_k^\top (\xi^{(k-1)-}) \right. \right.\nonumber\\
&\hspace{10mm}\left. \left. \hat P(\phi_k(\xi^{(k-1)-})) \Phi_k (\xi^{(k-1)-}) \overline{\partial \gamma}(s) \right)^{1/2} \right]\nonumber\\
& \le \frac{c_2}{c_1} \lambda^{k-k_0} \sqrt{\overline{\partial \gamma}^\top (s) \hat P(\overline{\gamma}(s)) \overline{\partial \gamma} (s)}, \; \forall k \in \bZ_{k_0+}
\label{IES:pf7}
\end{align}
for each~$(k_0, (x'_{k_0}, x''_{k_0}), \hat \xi^{(k_0-1)-}) \in \bZ \times (\bR^n \times \bR^n) \times \hat \Xi^{(k_0-1)-}$ and every $s \in \bR$, where in the left-hand side, the argument~$(k_0, \overline{\gamma} (s), \hat \xi^{(k_0-1)-})$ is dropped from~$\phi_k (\xi^{(k-1)-})$ and $\Phi_k (\xi^{(k-1)-})$. 

According to Remark~\ref{piecewise:rem}, $\phi_k$, $k \in \bZ_{k_0+}$ and $\Phi_k$, $k \in \bZ_{k_0+}$ are both piecewise continuous functions at each~$k_0 \in \bZ$. Note that a piecewise continuous function and stochastic process are both measurable, and the composition of measurable functions is again measurable~\cite[Proposition 2.1.1]{AL:06}. Therefore, in the left-hand side of~\eqref{IES:pf7}, $(\overline{\partial \gamma}^\top (s) \Phi_k^\top (\xi^{(k-1)-}) \hat P(\phi_k(\xi^{(k-1)-})) \Phi_k (\xi^{(k-1)-}) \overline{\partial \gamma}(s))^{1/2}$, $k \in \bZ_{k_0+}$ is $\cL$-measurable at each~$(k_0, (x_{k_0}, x'_{k_0})) \in \bZ \times (\bR^n \times \bR^n)$.

Taking the~$\mu$-integrations for both sides of~\eqref{IES:pf7} yield
\begin{align}
&\int_\bR \bE_0 \left[ \left( \overline{\partial \gamma}^\top (s) \Phi_k^\top (\xi^{(k-1)-}) \right. \right.\nonumber\\
&\hspace{12mm}\left. \left. \hat P(\phi_k(\xi^{(k-1)-})) \Phi_k (\xi^{(k-1)-}) \overline{\partial \gamma}(s) \right)^{1/2} \right] d\mu \nonumber\\
&\le \frac{c_2}{c_1} \lambda^{k-k_0} \int_\bR \sqrt{\overline{\partial \gamma}^\top (s) \hat P(\overline{\gamma}(s)) \overline{\partial \gamma} (s)} d\mu \nonumber\\
&= \frac{c_2}{c_1} \lambda^{k-k_0} \int_0^1 \sqrt{\frac{d^\top \gamma^*(s)}{ds} \hat P(\gamma^*(s)) \frac{d \gamma^*(s)}{ds}} ds \nonumber\\
&= \frac{c_2}{c_1} \lambda^{k-k_0} d_{\hat P} (x'_{k_0}, x''_{k_0}), \; \forall k \in \bZ_{k_0+}
\label{IES:pf8}
\end{align}
for each~$(k_0, (x'_{k_0}, x''_{k_0}), \hat \xi^{(k_0-1)-}) \in \bZ \times (\bR^n \times \bR^n) \times \hat \Xi^{(k_0-1)-}$, where the first equality follows from~\eqref{new_gamma} and the fact that the Lebesgue and Riemann integrals coincide with each other when $\sqrt{\frac{d^\top \gamma^*(s)}{ds} \hat P(\gamma^*(s)) \frac{d \gamma^*(s)}{ds}}$ is bounded and Riemann integrable with respect to~$s$ on~$[0,1]$ at each~$(x'_{k_0}, x''_{k_0}) \in \bR^n \times \bR^n$ (see e.g., \cite[Theorem 2.4.1]{AL:06}); the last equality follows from~\eqref{distance_upbd}. 

In~\eqref{IES:pf8}, the most right-hand side is bounded for each~$(k_0, (x'_{k_0}, x''_{k_0}),) \in \bZ \times (\bR^n \times \bR^n)$. This implies that the most left-hand side is~$\mu$-integrable at each~$(k_0, (x'_{k_0}, x''_{k_0}), \hat \xi^{(k_0-1)-}) \in \bZ \times (\bR^n \times \bR^n) \times \hat \Xi^{(k_0-1)-}$. Therefore, form the Fubini-Tonelli theorem~\cite[Section 5.2]{AL:06}, the order of the integrals in the most left-hand side is commutative; recall that $(\bR \times \Omega, \cL, \lambda)$ is complete and $\sigma$-finite. Namely, it follows that
\begin{align}
&\int_\bR \bE_0 \left[ \left( \overline{\partial \gamma}^\top (s) \Phi_k^\top (\xi^{(k-1)-}) \right. \right.\nonumber\\
&\hspace{12mm}\left. \left. \hat P(\phi_k(\xi^{(k-1)-})) \Phi_k (\xi^{(k-1)-}) \overline{\partial \gamma}(s) \right)^{1/2} \right] d\mu \nonumber\\
&= \bE_0 \biggl[ \int_\bR \left( \overline{\partial \gamma}^\top (s) \Phi_k^\top (\xi^{(k-1)-}) \right. \nonumber\\
&\hspace{15mm} \left. \hat P(\phi_k(\xi^{(k-1)-})) \Phi_k (\xi^{(k-1)-}) \overline{\partial \gamma}(s) \right)^{1/2} d\mu \biggr] \nonumber\\
&= \bE_0 \biggl[ \int_0^1 \left( \frac{\partial^\top \phi_k  (\xi^{(k-1)-})}{\partial s} \right. \nonumber\\
&\hspace{20mm} \left. \hat P(\phi_k(\xi^{(k-1)-})) \frac{\partial^\top \phi_k  (\xi^{(k-1)-})}{\partial s}\right)^{1/2} ds \biggr] \nonumber\\
&\ge \bE_0[ d_{\hat P} (\phi_k(\xi^{(k-1)-} ; k_0, x'_{k_0}, \hat \xi^{(k_0-1)-} ), \nonumber\\
&\hspace{18mm}\phi_k( \xi^{(k-1)-}; k_0, x''_{k_0},\hat \xi^{(k_0-1)-} ))], \; \forall k \in \bZ_{k_0+}
\label{IES:pf9}
\end{align}
for each~$(k_0, (x'_{k_0}, x''_{k_0}), \hat \xi^{(k_0-1)-}) \in \bZ \times (\bR^n \times \bR^n) \times \hat \Xi^{(k_0-1)-}$, where the first equality follows from the Fubini-Tonelli theorem; the second one follows from~\eqref{Phi_path},~\eqref{new_gamma}, and the fact that the Lebesgue and Riemann integrals coincide with each other by a similar reasoning as mentioned for~\eqref{IES:pf8}; the last inequality follows from~\eqref{distance:def} and the fact that $\phi_k(\xi^{(k-1)-} ; k_0, \gamma^*(s), \hat \xi^{(k_0-1)-} )$ is a path connecting $\phi_k( \xi^{(k-1)-} ; k_0, \gamma^* (0),\hat \xi^{(k_0-1)-} )$, $\gamma^* (0) = x'_{k_0}$ to $\phi_k(\xi^{(k-1)-}; k_0, \gamma^* (1), \hat \xi^{(k_0-1)-} )$, $\gamma^* (1) = x''_{k_0}$.

From~\eqref{IES:pf8} and~\eqref{IES:pf9}, we obtain
\begin{align*}
&\bE_0[ d_{\hat P} (\phi_k(\xi^{(k-1)-}; k_0, x'_{k_0}, \hat \xi^{(k_0-1)-} ),\\
&\hspace{20mm} \phi_k(\xi^{(k-1)-}; k_0, x''_{k_0}, \hat \xi^{(k_0-1)-} ))]\\
&\le \frac{c_2}{c_1} \lambda^{k - k_0}   d_{\hat P} (x'_{k_0}, x''_{k_0}), \; \forall k \in \bZ_{k_0+}
\end{align*}
for each~$(k_0, (x'_{k_0}, x''_{k_0})) \in \bZ \times (\bR^n \times \bR^n)$.
This implies that the system is UIES in the first moment.
\end{IEEEproof}
\section{Proof of Theorem~\ref{IES_Euclid:thm}}\label{IES_Euclid:app}

When $\hat P = I$, solving the corresponding Euler-Lagrange equation~\cite[Equation 5.3.2]{BCS:12} gives
\begin{align}
|x'-x''|^2 = \inf_{\gamma \in \Gamma (x',x'')} \int_0^1 \left| \frac{d \gamma (s)}{ds} \right|^2 ds,
\label{Euclid}
\end{align}
and the geodesic is the line segment~$\gamma^*(s) = (1-s) x'_{k_0} + s x''_{k_0}$. That is, when $\hat P = I$, we can directly use~\eqref{IES:pf4} for the sufficiency proof, but this is not true for general~$\hat P$. Utilizing~\eqref{Euclid}, we prove Theorem~\ref{IES_Euclid:thm} below.

\begin{IEEEproof}
(Sufficiency)
If $\hat P$ is identity,~\eqref{IES:pf4} reduces to 
\begin{align*}
\bE_0 \left[ |\delta x_k|^2 \right] \le \frac{c_2^2}{c_1^2}  \lambda^{2(k-k_0)} |\delta x_{k_0}|^2, \; \forall k \in \bZ_{k_0+}
\end{align*}
for each~$(k_0, (x_{k_0}, \delta x_{k_0}), \hat \xi^{(k_0-1)-}) \in \bZ \times (\bR^n \times \bR^n) \times \hat \Xi^{(k_0-1)-}$.
Substituting $(x_{k_0},\delta x_{k_0}) = (\gamma^* (s), d \gamma^* (s)/ds)$ with~$\gamma^*(s) = (1-s) x'_{k_0} + s x''_{k_0}$ (and consequently $d \gamma^*(s)/ds = x''_{k_0} - x'_{k_0}$) into this and taking the $\mu$-integration as in the proof of Theorem~\ref{IES:thm} yield
\begin{align*}
&\bE_0 \left[ \int_0^1 \left| \frac{\partial \phi_k (\xi^{k-}; k_0, \gamma^* (s), \hat \xi^{(k_0-1)-})}{\partial s} \right|^2  ds \right]\\
&= \bE_0 \left[\int_{\bR}  \left| \frac{\partial \phi_k (\xi^{k-}; k_0, \overline{\gamma} (s), \hat \xi^{(k_0-1)-})}{\partial s} \right|^2   d \mu \right] \\
&= \int_{\bR} \bE_0 \left[\left| \frac{\partial \phi_k (\xi^{k-}; k_0, \overline{\gamma} (s), \hat \xi^{(k_0-1)-})}{\partial s} \right|^2   \right] d \mu\\
& \le \frac{c_2^2}{c_1^2} \lambda^{2(k-k_0)}  |x'_{k_0} - x''_{k_0}|^2, \; \forall k \in \bZ_{k_0+}
\end{align*}
for each~$(k_0, (x'_{k_0}, x''_{k_0}), \hat \xi^{(k_0-1)-}) \in \bZ \times (\bR^n \times \bR^n) \times \hat \Xi^{(k_0-1)-}$, where~$\overline{\gamma}$ is defined in~\eqref{new_gamma}. From \eqref{Euclid} and the fact that $\phi_k(\xi^{(k-1)-} ; k_0, \gamma^* (s), \hat \xi^{(k_0-1)-} )$ is a path connecting $\phi_k( \xi^{(k-1)-} ; k_0, \gamma^* (0),\hat \xi^{(k_0-1)-} )$, $\gamma^* (0) = x'_{k_0}$ to $\phi_k(\xi^{(k-1)-}; k_0, \gamma^* (1), \hat \xi^{(k_0-1)-} )$, $\gamma^* (1) = x''_{k_0}$, the system is UIES in the second moment with respect to the Euclidean distance.

(Necessity)
(Step 1)
In fact, \eqref{Phi_path} holds for an arbitrary path~$\gamma \in \Gamma (x_{k_0}, x'_{k_0})$. As for the sufficiency proof, we choose~$\gamma (s) = (1-s) x_{k_0} + s x'_{k_0}$ (and thus~$d\gamma (s)/ds = x'_{k_0} - x_{k_0}$). Then, it follows from fundamental theorem of calculus, \eqref{sys_sol0}, and~\eqref{Phi_path} that 
\begin{align*}
&x'_k - x_k\\ 
&=\phi_k (\xi^{(k-1)-}; k_0, x'_{k_0}, \hat \xi^{(k_0-1)-}) \\
&\hspace{5mm} -  \phi_k (\xi^{(k-1)-}; k_0, x_{k_0}, \hat \xi^{(k_0-1)-}) \\
&= \int_0^1 \frac{\partial \phi_k (\xi^{(k-1)-}; k_0, (1-s) x_{k_0} + s x'_{k_0}, \hat \xi^{(k_0-1)-})}{\partial s} ds \\
&= \int_0^1 \Phi_k (\xi^{(k-1)-}; k_0, (1-s) x_{k_0} + s x'_{k_0}, \hat \xi^{(k_0-1)-})\\
&\hspace{40mm}  (x'_{k_0} - x_{k_0}) ds, \; \forall k \in \bZ_{k_0+}  
\end{align*}
for each~$(k_0, (x_{k_0}, x'_{k_0}), \hat \xi^{(k_0-1)-}) \in \bZ \times (\bR^n \times \bR^n) \times \hat \Xi^{(k_0-1)-}$. Substituting this into the definition~\eqref{IES:def} of the UIES in the second moment with respect to the Euclidean distance yields
\begin{align*}
&\bE_0 \Biggl[ \biggl| \int_0^1 \Phi_k (\xi^{(k-1)-}; k_0, (1-s) x_{k_0} + s x'_{k_0}, \hat \xi^{(k_0-1)-}) \\ 
&\hspace{8mm}  (x'_{k_0} - x_{k_0}) ds \biggr|^2 \Biggr] 
\le a^2 \lambda^{2(k-k_0)} | x'_{k_0} - x_{k_0} |^2, \\
&\hspace{60mm} \forall k \in \bZ_{k_0+}
\end{align*}
for each~$(k_0, (x_{k_0}, x'_{k_0}), \hat \xi^{(k_0-1)-}) \in \bZ \times (\bR^n \times \bR^n) \times \hat \Xi^{(k_0-1)-}$.
Since~$x'_{k_0} \in \bR^n$ is arbitrary, we choose~$x'_{k_0} = x_{k_0} + h v$ with $h \in \bR$ and $v \in \bR^n$. Substituting this yields
\begin{align*}
&\bE_0 \left[ \left| \int_0^1 \Phi_k (\xi^{(k-1)-}; k_0, x_{k_0} + s h v, \hat \xi^{(k_0-1)-}) v ds  \right|^2 \right]\\
&\le a^2 \lambda^{2(k-k_0)} | v|^2, \; \forall k \in \bZ_{k_0+},
\end{align*}
and the change of the variables~$\bar s = s h$ leads to 
\begin{align*}
&\bE_0 \left[ \left| \frac{1}{h} \int_0^h \Phi_k (\xi^{(k-1)-}; k_0, x_{k_0} + \bar s v, \hat \xi^{(k_0-1)-}) v  d \bar s  \right|^2 \right] \\
&\le a^2 \lambda^{2(k-k_0)} | v |^2, \; \forall k \in \bZ_{k_0+}
\end{align*}
for each~$(k_0, (x_{k_0}, x'_{k_0}), \hat \xi^{(k_0-1)-}) \in \bZ \times (\bR^n \times \bR^n) \times \hat \Xi^{(k_0-1)-}$.
Note that this holds for an arbitrary~$h \in \bR$, which implies
\begin{align}
&\liminf_{h \to 0}\bE_0 \left[ \left| \frac{1}{h} \int_0^h \Phi_k (\xi^{(k-1)-}; k_0, x_{k_0} + \bar s v, \hat \xi^{(k_0-1)-}) v  d \bar s  \right|^2 \right] \nonumber\\
&\le a^2 \lambda^{2(k-k_0)} | v |^2, \; \forall k \in \bZ_{k_0+}
\label{IES_Euclid:pf2}
\end{align}
for each~$(k_0, (x_{k_0}, v), \hat \xi^{(k_0-1)-}) \in \bZ \times (\bR^n \times \bR^n) \times \hat \Xi^{(k_0-1)-}$. Applying Fatou's lemma~\cite[Theorem 2.3.7]{AL:06} to the left-hand side yields
\begin{align}
&\bE_0 \left[ \liminf_{h \to 0} \left| \frac{1}{h} \int_0^h \Phi_k (\xi^{(k-1)-}) v  d \bar s  \right|^2 \right] \nonumber\\
&\le \liminf_{h \to 0} \bE_0 \left[ \left| \frac{1}{h} \int_0^h \Phi_k (\xi^{(k-1)-}) v  d \bar s  \right|^2 \right], \; \forall k \in \bZ_{k_0+}
\label{IES_Euclid:pf3}
\end{align}
for each~$(k_0, (x_{k_0}, v), \hat \xi^{(k_0-1)-}) \in \bZ \times (\bR^n \times \bR^n) \times \hat \Xi^{(k_0-1)-}$.

We consider the left-hand side of~\eqref{IES_Euclid:pf3}. Here, we take $\phi_k$ and $\Phi_k$ as the mappings $\phi_k(x_{k_0}, \eta^{(k-1)-}; k_0)$ and~$\Phi_k(x_{k_0}, \eta^{(k-1)-}; k_0)$; recall Remark~\ref{mapping:rem}. Applying product and sum rules of the limit and fundamental theorem of calculus in order lead to
\begin{align}
&\lim_{h \to 0} \left| \frac{1}{h} \int_0^h \Phi_k (x_{k_0} + \bar s v, \eta^{(k-1)-}; k_0) v  d \bar s  \right|^2 \nonumber\\
&= \left| \lim_{h \to 0} \frac{1}{h} \int_0^h \Phi_k (x_{k_0} + \bar s v, \eta^{(k-1)-}; k_0) v  d \bar s  \right|^2 \nonumber\\
&= \left| \Phi_k (x_{k_0}, \eta^{(k-1)-}; k_0) v \right|^2, \; \forall k \in \bZ_{k_0+}
\label{IES_Euclid:pf4}
\end{align}
for each $(k_0, (x_{k_0}, v), \eta^{(k-1)-}) \in \bZ \times (\bR^n \times \bR^n) \times (\bR^m)^{\bZ_{[k_0,k-1]}}$. These equalities imply the existence of the limit in the most left-hand side. Therefore, the limit inferior of the left-hand side of~\eqref{IES_Euclid:pf3} is equivalent to the limit. Combining~\eqref{IES_Euclid:pf2} -- \eqref{IES_Euclid:pf4} leads to, for stochastic processes,
\begin{align*}
\bE_0 \left[\left| \Phi_k (\xi^{(k-1)-}; k_0, x_{k_0}, \hat \xi^{(k_0-1)-}) v \right|^2 \right] \le a^2 \lambda^{2(k-k_0)} | v |^2, \\
 \forall k \in \bZ_{k_0+}
\end{align*}
for each $(k_0, x_{k_0}, \hat \xi^{(k_0-1)-}) \in \bZ \times \bR^n \times \hat \Xi^{(k_0-1)-}$ and $v \in \bR^n$. Since~$v \in \bR^n$ is arbitrary, it holds that
\begin{align}
&\bE_0 \left[\sigma_{\max}^2 \left( \Phi_k (\xi^{(k-1)-}; k_0, x_{k_0}, \hat \xi^{(k_0-1)-})  \right) \right] \le a^2 \lambda^{2(k-k_0)}, \nonumber\\
&\hspace{60mm}\forall k \in \bZ_{k_0+} \label{IES_Euclid:pf5}
\end{align}
for each $(k_0, x_{k_0}, \hat \xi^{(k_0-1)-}) \in \bZ \times \bR^n \times \hat \Xi^{(k_0-1)-}$, where~$\sigma_{\max}( \cdot )$ denotes the largest singular value.

(Step 2)
Here, we again consider $\phi_k$ and $\Phi_k$ as the mappings $\phi_k(x_{k_0}, \eta^{(k-1)-}; k_0)$ and~$\Phi_k(x_{k_0}, \eta^{(k-1)-}; k_0)$; recall Remark~\ref{mapping:rem}. Let us take~$\lambda_1$ such that~$\lambda < \lambda_1 <1$, and define the $K$-dependent matrix-valued mapping $P_K: \bZ \times \bR^n \times (\bR^m)^{\bZ_{k_0+}} \to \bS_{\succ 0}^{n\times n}$, $K \in \bZ_{k_0+}$ such that
\begin{align}
&P_K(k_0, x_{k_0}, S_{k_0} \eta^{k_0+}) \nonumber\\
&:=\frac{1}{\lambda_1^2}\sum_{k=k_0}^K \frac{1}{\lambda_1^{2(k-k_0)}} \Phi_k^\top  (x_{k_0}, \eta^{(k-1)-}; k_0) \nonumber\\
&\hspace{30mm}\Phi_k  (x_{k_0}, \eta^{(k-1)-}; k_0), \; K \in \bZ_{k_0+} \label{PK}
\end{align}
for each~$(k_0, x_{k_0}, \eta^{k_0+}) \in \bZ \times \bR^n \times (\bR^m)^{\bZ_{k_0+}}$. Note that $(P_K (k_0, x_{k_0}, S_{k_0} \eta^{k_0+}) )_{K\in \bZ_{k_0+} }$ is an increasing sequence with respect to the relation~$\preceq$ for each~$(k_0, x_{k_0}, \eta^{k_0+}) \in \bZ \times \bR^n \times (\bR^m)^{\bZ_{k_0+}}$, and thus $P_K (k_0, x_{k_0}, S_{k_0} \eta^{k_0+})$ has the (pointwise convergence) limit:
\begin{align}
&P (k_0, x_{k_0}, S_{k_0} \eta^{k_0+})\nonumber\\
& := \lim_{K \to \infty}  P_K (k_0, x_{k_0}, S_{k_0} \eta^{k_0+}) \in [-\infty, \infty]^{n\times n}
\label{P}
\end{align}
for each~$(k_0, x_{k_0}, \eta^{k_0+}) \in \bZ \times \bR^n \times (\bR^m)^{\bZ_{k_0+}}$.

Now, we consider stochastic processes.
According to Remark~\ref{piecewise:rem}, $\phi_k(\xi^{(k-1)-})$, $k \in \bZ_{k_0+}$ and $\Phi_k(\xi^{(k-1)-})$, $k \in \bZ_{k_0+}$ are both $\cF_{k-1}$-measurable for each~$(k_0, x_{k_0}, \hat \xi^{(k_0-1)-}) \in \bZ \times \bR^n \times \hat \Xi^{(k_0-1)-}$, and thus $P_K (k_0, x_{k_0}, S_{k_0} \xi^{k_0+}) $, $K \in \bZ_{k_0+}$ is $\cF$-measurable for each $(k_0, x_{k_0}, \hat \xi^{(k_0-1)-}) \in \bZ \times \bR^n \times \hat \Xi^{(k_0-1)-}$. Since the limit of a sequence of measurable functions is again measurable \cite[Proposition 2.1.5]{AL:06}, $P(k_0, x_{k_0}, S_{k_0} \xi^{k_0+})$ is $\cF$-measurable for each~$(k_0, x_{k_0}, \hat \xi^{(k_0-1)-}) \in \bZ \times \bR^n \times \hat \Xi^{(k_0-1)-}$. Therefore, the monotone convergence theorem \cite[Theorem 2.3.4]{AL:06} is applicable for $P(k_0, x_{k_0}, S_{k_0} \xi^{k_0+})$, namely
\begin{align}
\bE_0 \left[ P(k_0, x_{k_0}, S_{k_0} \xi^{k_0+}) \right] = \lim_{K \to \infty} \bE_0\left[ P_K(k_0, x_{k_0}, S_{k_0} \xi^{k_0+}) \right]
\label{IES_Euclid:pf6}
\end{align}
for each~$(k_0, x_{k_0}, \hat \xi^{(k_0-1)-}) \in \bZ \times \bR^n \times \hat \Xi^{(k_0-1)-}$. Note that these conditional expectations are not necessarily to be finite.

(Step 3)
From~\eqref{Phi2} and~\eqref{PK}, it follows that 
\begin{align*}
\frac{1}{\lambda_1^2} I \preceq P_K(k_0, x_{k_0}, S_{k_0} \xi^{k_0+}) \mbox{ a.s.}, \; \forall K \in \bZ_{k_0+} 
\end{align*}
for each~$(k_0, x_{k_0}, \hat \xi^{(k_0-1)-}) \in \bZ \times \bR^n \times \hat \Xi^{(k_0-1)-}$. This inequality is preserved under taking the conditional expectations~$\bE_0[\cdot ]$ of both sides. From~\eqref{IES_Euclid:pf5} and~\eqref{PK}, the conditional expectation~$\bE_0 [P_K(k_0, x_{k_0}, S_{k_0} \xi^{k_0+})]$ is also upper bounded. That is, it holds that
\begin{align}
c_1^2 I \le \bE_0 [P_K(k_0, x_{k_0}, S_{k_0} \xi^{k_0+})] \le c_2^2 I, \; \forall K \in \bZ_{k_0+}
\label{IES_Euclid:pf7}
\end{align}
for each~$(k_0, x_{k_0}, \hat \xi^{(k_0-1)-}) \in \bZ \times \bR^n \times \hat \Xi^{(k_0-1)-}$, where~$c_1:= 1/\lambda_1$ and
\begin{align*}
c_2:= \frac{a}{\lambda_1} \lim_{K \to \infty} \sum_{k=k_0}^K \left(\frac{\lambda} {\lambda_1}\right)^{k-k_0} > 0.
\end{align*}
Note that $c_2$ is a $k_0$-independent positive constant because of $\lambda < \lambda_1$.
Since \eqref{IES_Euclid:pf7} holds for an arbitrary $K \in \bZ_{k_0+}$, taking $K \to \infty$ and using \eqref{IES_Euclid:pf6} yield
\begin{align*}
c_1^2 I \le \bE_0 \left[ P(k_0, x_{k_0}, S_{k_0} \xi^{k_0+}) \right] \le c_2^2 I, \; \forall K \in \bZ_{k_0+}
\end{align*}
for each~$(k_0, x_{k_0}, \hat \xi^{(k_0-1)-}) \in \bZ \times \bR^n \times \hat \Xi^{(k_0-1)-}$. This is nothing but~\eqref{Pbd_Euclid:cond}. 

(Step 4) 
Here, we again consider $\phi_k$ and $\Phi_k$ as the mappings $\phi_k(x_{k_0}, \eta^{(k-1)-}; k_0)$ and~$\Phi_k(x_{k_0}, \eta^{(k-1)-}; k_0)$; recall Remark~\ref{mapping:rem}. From~\eqref{PK}, $P_K(k_0, x_{k_0}, S_{k_0} \eta^{k_0+})$ satisfies
\begin{align}
&\lambda_1^2 P_K(k_0, x_{k_0}, S_{k_0} \eta^{k_0+}) \nonumber\\
&-\frac{\partial^\top f_{k_0} (x_{k_0}, \eta_{k_0})}{\partial x_{k_0}} \nonumber\\
&\hspace{5mm}P_K(k_0+1, f_{k_0}(x_{k_0},\eta_{k_0}), S_{k_0+1} \eta^{(k_0+1)+}; \eta_{k_0}) \nonumber\\
&\hspace{5mm}\frac{\partial f_{k_0} (x_{k_0}, \eta_{k_0})}{\partial x_{k_0}} = I, \; \forall K \in \bZ_{k_0+}
\label{IES_Euclid:pf8}
\end{align}
for each~$(k_0, x_{k_0}, \eta^{k_0+}) \in \bZ \times \bR^n \times (\bR^m)^{\bZ_{k_0+}}$, where the equality
\begin{align*}
\Phi_{k_0+1} (x_{k_0}, \eta_{k_0}; k_0) 
 = \frac{\partial f_{k_0} (x_{k_0}, \eta_{k_0})}{\partial x_{k_0}}
\end{align*}
following from~\eqref{sys_sol_eta},~\eqref{Phi3_eta}, and $\phi_{k_0} = x_{k_0}$ is used.
Note that $P_K$ in the second term of the left-hand side of~\eqref{IES_Euclid:pf8} depends on $\eta_{k_0}$. 

From~\eqref{IES_Euclid:pf8}, the corresponding stochastic processes satisfy
\begin{align*}
&\frac{\partial^\top f_{k_0} (x_{k_0}, \xi_{k_0})}{\partial x_{k_0}} P_K(k_0+1, f_{k_0}(x_{k_0},\xi_{k_0}), S_{k_0+1} \xi^{(k_0+1)+})\\
&\frac{\partial f_{k_0} (x_{k_0}, \xi_{k_0})}{\partial x_{k_0}}
 \preceq \lambda_1^2 P_K(k_0, x_{k_0}, S_{k_0} \xi^{k_0+}), \;
 \forall K \in \bZ_{k_0+}
\end{align*}
for each~$(k_0, x_{k_0}, \hat \xi^{(k_0-1)-}) \in \bZ \times \bR^n \times \hat \Xi^{(k_0-1)-}$. Taking the conditional expectations of~$\bE_0[ \cdot | \cF_{k_0}]$ and~$\bE_0[ \cdot ]$ in order for both sides and using the fact that~$(\cF_k)_{k \in \bZ_{k_0+}}$ is a filtration on $(\Omega, \cF, \bP)$ for each~$\hat \xi^{(k_0-1)-} \in \hat \Xi^{(k_0-1)-}$ for the right-hand side, it follows that 
\begin{align*}
&E_0 \left[ \frac{\partial^\top f_{k_0} (x_{k_0}, \xi_{k_0})}{\partial x_{k_0}} \right.\\
&\hspace{7mm}\bE_0[P_K(k_0+1, f_{k_0}(x_{k_0},\xi_{k_0}), S_{k_0+1} \xi^{(k_0+1)+}) | \cF_{k_0}]\\
&\hspace{7mm}\left. \frac{\partial f_{k_0} (x_{k_0}, \xi_{k_0})}{\partial x_{k_0}} \right] \preceq \lambda_1^2 \bE_0[ P_K(k_0, x_{k_0}, S_{k_0} \xi^{k_0+})], \\
&\hspace{60mm} \forall K \in \bZ_{k_0+}
\end{align*}
for each~$(k_0, x_{k_0}, \hat \xi^{(k_0-1)-}) \in \bZ \times \bR^n \times \hat \Xi^{(k_0-1)-}$. Finally, taking~$K \to \infty$ with~\eqref{IES_Euclid:pf6} yields~\eqref{IES:cond}. 
\end{IEEEproof}

\bibliographystyle{IEEEtran} 
\bibliography{Ref}

\end{document}